\documentclass{article}
\usepackage[utf8]{inputenc}
\usepackage{amsmath}
\usepackage{amssymb}
\usepackage[margin=1.2in]{geometry}
\usepackage{bbold, paralist}
\usepackage{multirow}
\usepackage{makecell}
\usepackage{graphicx}
\newcommand{\R}{\mathbb{R}}
\usepackage{xcolor}
\usepackage{multicol}
\usepackage{setspace}

\DeclareMathOperator{\Tr}{Tr}
\DeclareMathOperator*{\argmax}{arg\,max}

\DeclareMathOperator{\gammab}{\boldsymbol{\gamma}}

\DeclareMathOperator{\betab}{\boldsymbol{\beta}}

\DeclareMathOperator{\No}{\mathcal{N}}

\newcommand\independent{\protect\mathpalette{\protect\independenT}{\perp}}
\def\independenT#1#2{\mathrel{\rlap{$#1#2$}\mkern2mu{#1#2}}}

\title{\textbf{Combined shrinkage of fixed and random effects in linear mixed models using empirical Bayes.}}
\author{}
\author{Matteo Amestoy,
R. Vermeulen,
 Mark A. van de Wiel,
 Wessel N. van Wieringen}

\usepackage[backend=biber,style = apa]{biblatex} %Imports biblatex package
\begin{document}

\maketitle

\vspace{1cm}
\vspace*{0.5cm}\textbf{Abstract:}

% We propose a novel data-driven method for the joint selection of the fixed and random effect priors' parameters of a linear mixed model (LMM). It facilitates the estimation of LMMs with complex random effect structures from possibly high-dimensional data. We do so using a Bayesian solver. That requires informed choices for the prior parameters. This need not be readily available. In particular, for the covariance of the random effects.

% Our proposed empirical Bayes method allows for the automatic selection of the priors' parameters. The method involves maximizing the model's marginal likelihood, which is computed efficiently through the utilization of a Laplace approximation. In simulations we show that our method significantly improves the accuracy of parameter estimates and enhances the model's predictive power. Furthermore, its application to a real-world air pollution and health dataset demonstrates the ability to enhance predictive accuracy by enabling the use of a more complex model.
A novel data-driven methodology is presented for the joint selection of prior parameters for both fixed and random effects in Linear Mixed Models (LMMs). This approach facilitates the estimation of complex random-effects structures, as well as potentially high-dimensional data. Although Bayesian frameworks require the specification of informative prior parameters, such values are often unavailable \textit{a priori}---especially for random-effect covariances. The proposed method automates this selection through an Empirical Bayes framework, which maximizes the marginal likelihood using an efficient Laplace approximation. Numerical simulations demonstrate that this methodology significantly enhances parameter estimation accuracy and predictive performance. Finally, an application to a real-world air pollution and health dataset illustrates how the method enables the use of more sophisticated and statistically appropriate models to improve predictive outcomes.

\clearpage
\section{Introduction}
Linear Mixed Models (LMMs) constitute a flexible and robust statistical framework for the analysis of continuous outcomes characterized by hierarchical or clustered correlation structures. Due to their versatility, LMMs have become the predominant methodology for longitudinal data analysis, wherein repeated measurements are obtained from the same experimental units over time. These models are extensively employed in epidemiological research to investigate associations between risk factors and health outcomes while accounting for the inherent nested structure of the data.

LMMs simultaneously characterize both the mean and the variance of the response variable. Fixed-effect parameters represent population-level associations between the outcome and explanatory variables, whereas random-effect parameters and residual variance components capture individual-level variability and within-subject correlation. This approach facilitates a nuanced understanding of the factors influencing the outcome while accounting for the structure of the data.

A contemporary trend in epidemiological studies involves the acquisition of increasingly expansive datasets, often facilitated by high-throughput technologies pertaining to both participants and their environments. The emergence of such complex data frequently presents scenarios where conventional statistical methods yield unreliable estimates. This typically occurs when the dimensionality of the parameter space exceeds the effective sample size.

While the estimation of fixed effects in LMMs can be hampered by high-dimensionality—a phenomenon analogous to the $p > n$ problem in standard linear regression—numerous regularized solutions have been developed to address this. In contrast, the reliable estimation of random-effect parameters relative to study design has received comparatively less scholarly attention. This disparity is notable, as model specification in practical applications is often constrained by the estimability of these random effects. This paper identifies two common structural configurations that impede estimation:
\begin{compactitem}
    \item[$\circ$] the number of covariates in the random effects model exceeds the number of statistical units.
    \item[$\circ$] the number of covariates in the random effects model exceeds the average number of repeated measurements per statistical unit.
\end{compactitem}
The dataset provided by \cite{van_nunen_short-term_2021} serves as the motivating example for this study, instantiating the two problematic structures identified above. This dataset stems from a three-wave longitudinal study conducted across three distinct locations, encompassing measurements of individual characteristics and environmental exposures. The primary objective was to quantify the impact of these exposures on specific health outcomes. A standard LMM approach would necessitate two levels of random effects: one for individuals and one for locations.

For the location-level random effect, the study includes only three statistical units, which precludes a stable estimation of the variance attributable to location. At the individual level, the maximum of three observations per participant limits the number of random-effect covariates the model can accommodate. Specifically, including more than three random variables at the individual level (e.g., a random intercept and two random slopes) using standard software such as the R package lme4 results in an error indicating that the number of observations is less than or equal to the number of random effects, suggesting the parameters may be unidentifiable. However, while the model remains theoretically identifiable \parencite{amestoy_identifiability_2024}, the resulting estimates are often numerically unstable and unreliable.

In the original analysis of the motivating dataset, the authors addressed these estimation challenges through ad hoc methodologies, fitting separate models for each location and aggregating results via meta-analysis, while restricting the individual-level random effect to a single intercept. Conversely, an integrated LMM including both location and individual random effects can be estimated using regularization. Furthermore, although these are not present in our illustrative examplre, regularization can accommodate high-dimensional fixed effects.

Although regularization of fixed effects in LMMs has garnered significant attention within the literature \parencite{an_shrinkage_2009,groll_variable_2014,yang_bayesian_2020}, the relevance of regularizing the random effects variance parameter remains largely overlooked. Bayesian methods inherently regularize both parameters by incorporating prior knowledge into the model. However, selecting such informed priors is often infeasible, particularly for the variance of random effects. Although certain methods have been proposed for establishing the prior of random effects parameters \parencite{kass_default_2006,akinc_bayesian_2018}, their intricate and technical nature often forces practitioners to simplify their models.

This paper introduces a data-driven empirical Bayes method for the simultaneous selection of optimal priors for both fixed and random-effect parameters. This is achieved by maximizing a Laplace approximation of the marginal likelihood. We demonstrate that imprecise estimation of random effects can adversely impact the model’s predictive performance and the estimation of critical quantities, such as the variance of fixed-effect estimates. This remains true even when random-effect estimation has a negligible impact on the point estimates of fixed effects.

While it is recognized that fixed-effect estimates are often robust to the misspecification of random-effect distributions \parencite{verbeke_effect_1997, taylor_does_1998}, the proposed procedure ensures reliable estimation of both the fixed effects and the covariance matrix of the random effects. This dual reliability enhances the downstream utility of the LMM in two critical ways:
\begin{itemize}
    \item Inferential Precision: Inference regarding fixed effects is improved through more accurate estimates of their variance. This is particularly vital for significance testing and the calculation of $p$-values, as implemented in the lmerTest package \cite{kuznetsova_lmertest_2017}.
    \item Predictive Accuracy: Conditional predictions benefit from precise variance estimates, as the LMM relies on these to account for correlations among observations. The propagation of accurate covariance estimates through the model significantly improves overall predictive performance.
\end{itemize}

The remainder of this paper is organized as follows. Section \ref{sec:method} details the proposed methodology, beginning with the formal LMM specification followed by the empirical Bayes approach for prior selection. Section \ref{sec:data} evaluates the method using simulated data and the empirical dataset from \cite{van_nunen_short-term_2021}.

\section{Empirical Bayes regularised LMM }\label{sec:method}
Section \ref{sec:method} details the methodological framework for a fully data-driven Bayesian estimation of LMMs. This approach utilizes an empirical Bayes (EB) strategy to determine the prior distribution. The following sections define the LMM structure and the specified priors, and the subsequent EB estimation procedure.

Within the Bayesian framework, the parameter vector $\boldsymbol{\theta} \in \mathbb{R}^p$ is treated as a random variable. The objective is to characterize the posterior distribution of $\boldsymbol{\theta}$ given the observed data $\mathbf{Y}$. According to Bayes' Theorem:$$p(\boldsymbol{\theta}|\mathbf{Y}) \propto p(\mathbf{Y}|\boldsymbol{\theta}) \, p(\boldsymbol{\theta}|\boldsymbol{\Theta}),$$where $p(\mathbf{Y}|\boldsymbol{\theta})$ represents the model likelihood and $p(\boldsymbol{\theta}|\boldsymbol{\Theta})$ denotes the prior distribution, governed by the hyperparameter $\boldsymbol{\Theta}$. While hyperparameter selection may be informed by domain expertise, such information is frequently unavailable, necessitating a data-driven approach.

\subsection{Linear mixed model}\label{subsec:LMM}
Here we describe the LMM to clarify the observations' covariance matrix structure and its dependence on the parameters and design particulars, such as the sample size and number of repetitions. We then give the parameters priors to obtain a Bayesian linear mixed model formulation. 

We assume a study involving $R \in \mathbb{N}$ statistical units, e.g. individuals and hospitals if $R=2$, that have been characterized repeatedly concerning an outcome. We denote the total number of observations by $N  \in \mathbb{N}$. For each observation $i = 1, \ldots, N$, a continuous response variable $Y_i \in \mathbb{R}$ is observed. The latter is accompanied by two sets of explanatory variables related to the linear mixed model's fixed and random effect. The first set comprises covariate information relating the fixed effect part and  is the $p$ dimensional row vector $\mathbf{X}_{i, \ast}$. The second set, corresponding to the random effect part, are the $q_r$-dimensional row vectors $(\mathbf{U}_{r})_{i, \ast}$ for the statistical units $r=1, \ldots, R$, e.g. $q_r=2$ for a random slope and a random intercept for unit $r$. The covariate information relating to the random effect part is often reformulated to simplify the likelihood formulation. Hereto we define the matrix $\mathbf{Z} := ( \mathbf{K}_1 \bullet \mathbf{U}_1, \ldots, \mathbf{K}_R \bullet \mathbf{U}_R)$, where $\bullet$ is face splitting operator. The $\mathbf{K}_r$ are binary matrices of dimensional $N \times m_r$, where $m_r$ is the number of levels in the statistical unit $r$. Each row of the $\mathbf{K}_r$ comprises a single entry equalling one and all others equalling zero. We denote by $k_r(i)$ the position of the 1 in the $i$-th row of matrix $\mathbf{K}_r$. For instance, the $\mathbf{Z}_{i, \ast}$ may then comprise covariates indicating to which statistical unit the observation $i$ belongs. Finally, let $\mathbf{Y}$ be the vector of all observations and the covariate matrices $\mathbf{X}$ be formed by stacking their row counterparts of all observations.

We present two formulations of the linear mixed model. The first formulation, used by most software implementations of the linear mixed model solvers, e.g. \texttt{lmer} \parencite{bates_fitting_2015}, is:
\begin{equation}\label{eq:indLevel}
Y_i = \mathbf{X}_{i,\ast} \betab+\sum\nolimits_{r=1}^R (\mathbf{U}_{r})_{i,\ast} \gammab_{r,k_r(i)} +\varepsilon_i. 
\end{equation}
In the above, $\betab\in \R^p$ is the fixed effect parameter. The $\gammab_{r,k_r(i)}$ are the $q_r$-dimensional random effects distributed as $\gammab_{r,k_r(i)} \sim \No(\boldsymbol{0}_{q_r}, \mathbf{\Sigma}_r)$ with covariance matrices $\mathbf{\Sigma}_r$. The random effects $\gammab_{r,k_r(i)}$ and $\gammab_{r',k_{r'(i)}}$ are assumed to be independent for $r \not= r'$ and, similarly, $\gammab_{r,k_r(i)} \independent \gammab_{r,k_{r(i')}}$ if $k_{r(i')} \not= k_{r(i')}$. Finally, the model includes an error term $\varepsilon_i$ that follows the normal law $\No(\boldsymbol{0},\sigma^2)$ with the $\varepsilon_i$ independent between different observations. The (unknown) parameters of the model thus are $\boldsymbol{\theta} = \{ \betab, \mathbf{\Sigma}_1,\cdots,\mathbf{\Sigma}_R, \sigma^2 \}$.

The alternative formulation of the LMM that is mainly used to simplify the likelihood formulation, and subsequent derivations of model properties and those of its estimator. This formulation is at the population level and therefore in matrix form:
\begin{eqnarray}\label{eq:popLevel}
\mathbf{Y} & = & \mathbf{X}\betab+ \mathbf{Z} \boldsymbol{\gamma} +\boldsymbol{\varepsilon}
\, \, \, = \, \, \,  \mathbf{X}\betab+\sum\nolimits_{r=1}^R \mathbf{Z}_r \gammab_r +\boldsymbol{\varepsilon},
\end{eqnarray}
where $\boldsymbol{\gamma} = (\boldsymbol{\gamma}_1^{\top}, \ldots, \boldsymbol{\gamma}_R^{\top})^{\top}$ with the $r$-th random effect $\gammab_r = (\gammab_{r,1}^\top,\cdots,\gammab_{r,m_r}^\top)^\top \in \mathbb{R}^{m_r q_r}$ and $\mathbf{Z}_r = \mathbf{K}_r\bullet \mathbf{U}_r$. The random effects are distributed as $\gammab_r\sim\No(\boldsymbol{0}_{m_rq_r},\mathbf{W}_r)$ with block-diagonal covariance matrix $\mathbf{W}_r$, composed of $m_r$ repeated blocks of $\boldsymbol{\Sigma}_r$, i.e. $\mathbf{W}_r=\mathbf{I}_{m_r}\otimes \boldsymbol{\Sigma}_r$. Finally,  we have $\boldsymbol{\varepsilon}\sim \No(\boldsymbol{0}_n,\sigma^2 \mathbf{I}_{n})$ for the error vector. This alternative formulation of the LMM shows that $\mathbf{Y}\sim \mathcal{N} (\mathbf{X}\betab,\sigma^2 \mathbf{I}_n+\sum_{r=1}^{R} \mathbf{Z}_r \mathbf{W}_r \mathbf{Z}_r^\top)$. With this at hand, the likelihood can be straightforwardly derived. 

To complete the Bayesian setup of the LMM, we use conditional conjugate priors as they are the natural default priors and allow for easier and faster maximum a posteriori computation:
\begin{eqnarray}
\boldsymbol{\beta} \sim  \mathcal{N}(0_p, \sigma^2 \lambda^{-1} \mathbf{I}_p),\,\,
    \mathbf{\Sigma}_r \sim   \mathcal{W}^{-1}(\eta_r,\boldsymbol{\Phi}_r),\,\,
    \sigma^2 \sim  \mathcal{IG}(\alpha_1,\alpha_2).
\end{eqnarray}
The Gaussian prior on $\boldsymbol{\beta}$ is equivalent to a ridge penalty, where the scalar parameter $\lambda$ controls for the intensity of the regularisation. The prior distribution on the error variance parameter $\sigma$ is set to a non-informative one with $\alpha_1=\alpha_2=\alpha$ set to a small value \parencite{gelman_prior_2006}. We use an inverse-Wishart prior of degree of freedom $\eta_r$ and scale matrix $\boldsymbol{\Phi}_r$, for the random effect covariance matrices. The parameters of the priors, or hyperparameters, thus are $\boldsymbol{\Theta} = \{ \lambda,\eta_1,\boldsymbol{\Phi}_1,\cdots,\eta_R,\boldsymbol{\Phi}_R\} $ and, in the next section, we present a data-driven method to define them.

We make two remarks regarding the priors. Firstly, an alternative to this prior on the covariance matrices would be to put a Wishart prior on the precision matrices, $\mathbf{\Sigma}_r^{-1} \sim   \mathcal{W}(\eta'_r,\boldsymbol{\Phi}'_r)$, which is sometimes preferred in the Bayesian literature. Both priors are equivalent in a Bayesian framework as they lead to the same posterior law with $\eta_r =\eta_r'$ and $\boldsymbol{\Phi}^{-1}_r=\boldsymbol{\Phi}'_r$. For computational reasons we prefer an inverse Wishart prior. Secondly, the interaction between the scale matrix and degree of freedom parameter is complex, as both control for the prior mode's location and distribution's dispersion. In the next section we provide a more intuitive parametrisation for the case at hand.

\subsection{Empirical Bayes prior choice and MAP estimation}
This section presents an EB approach to select the hyperparameters of the prior distribution. The EB methodology identifies these hyperparameters from the observed data in an informed manner by maximizing the marginal likelihood, denoted as $\text{ML}(\cdot)$. The marginal likelihood represents the likelihood of the data given the hyperparameters after the model parameters have been integrated out \parencite{van_de_wiel_learning_2019}. Formally, the optimal hyperparameter set $\boldsymbol{\Theta}^*$ is defined as:
\begin{eqnarray} \label{form:hyperMLdef}
\boldsymbol{\Theta}^* & = & \argmax_{\boldsymbol{\Theta}} \text{ML}(\boldsymbol{\Theta}) \, \, \, = \, \, \, \argmax_{\boldsymbol{\Theta}}\int f(\mathbf{Y} \, | \, \boldsymbol{\theta}) \, \pi(\boldsymbol{\theta} \, | \, \boldsymbol{\Theta}) \, d\boldsymbol{\theta}, 
\end{eqnarray}
where $f(\cdot)$ is the likelihood of the data given the parameters and $\pi(\cdot)$ is the prior.

To the best of our knowledge, the marginal likelihood of the LMM lacks a closed-form solution and requires approximation. Potential methods include Markov Chain Monte Carlo (MCMC) (\cite{roy_selection_2017}) or direct posterior approximation via Integrated Nested Laplace Approximation (\texttt{INLA}) \parencite{gomez-rubio_bayesian_2020}. However, the computational requirements of these methods often do not scale efficiently with the number of model parameters. Furthermore, the maximization of the marginal likelihood over the hyperparameter space necessitates multiple evaluations, further increasing the requirement for computational efficiency.

\begin{figure}[ht!]
    \centering
    \includegraphics[scale = 0.3]{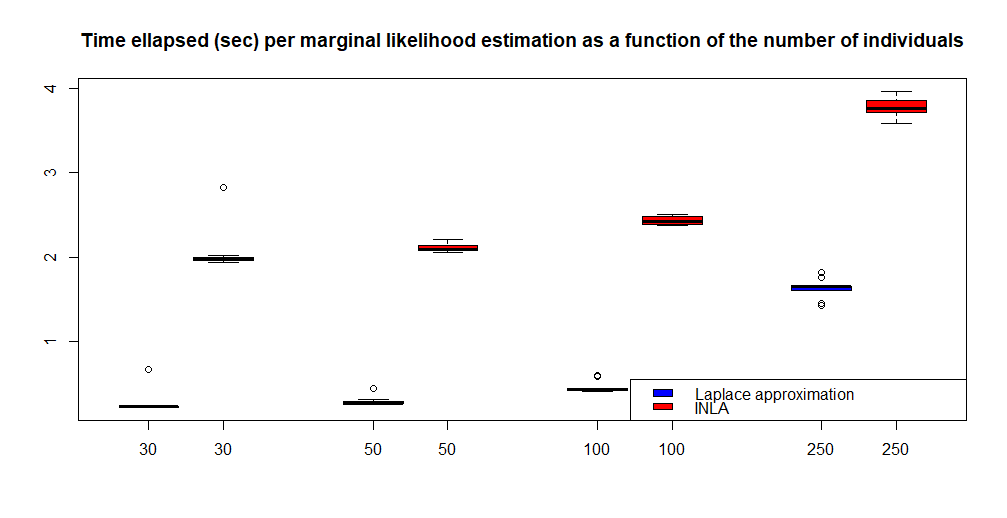}
    \caption{Computational time (in seconds) for marginal likelihood estimation of a LMM with a two-dimensional fixed effect and a four-dimensional individual-level random effect. The x-axis represents the number of individuals in the dataset, with each individual having three repeated observations. Each experiment is repeated ten times. Results are shown for the Laplace approximation (blue) and INLA (red).}
    \label{fig:Inla}
\end{figure}

To address this, the marginal likelihood is approximated using a Laplace approximation. Let $\log[L(\boldsymbol{\theta};\mathbf{Y},\boldsymbol{\Theta})] = \log[f(\mathbf{Y} \, | \,\boldsymbol{\theta})] + \log[\pi(\boldsymbol{\theta} \, | \, \boldsymbol{\Theta})]$ represent the logarithm of the integrand. Under relatively mild assumptions \parencite{shun_laplace_1995}:
\begin{eqnarray*}
\text{ML}(\boldsymbol{\Theta}) & = & \int \exp\{ \log[L(\boldsymbol{\theta};\mathbf{Y},\boldsymbol{\Theta})] \} \, d \boldsymbol{\theta} \, \, \, \simeq \, \, \, (2\pi)^{d/2} \frac{\exp \{ \log[L(\boldsymbol{\theta}^\ast;\mathbf{Y},\boldsymbol{\Theta})] \} }{ \{ \mbox{det}[ -H(\boldsymbol{\theta}^\ast)]\}^{1/2}}, 
\end{eqnarray*}
where $\boldsymbol{\theta}^*$ is the Maximum A Posteriori (MAP) estimator of $\boldsymbol{\theta}$ and $H(\cdot)$ denotes the Hessian matrix of $\log[L(\boldsymbol{\theta};\mathbf{Y},\boldsymbol{\Theta})]$ with respect to the model parameters. While the Hessian matrix possesses a closed-form expression (see Appendix), $\boldsymbol{\theta}^*$ must be computed numerically. By employing this Laplace approximation, it is sufficient to determine $\boldsymbol{\theta}^*$ rather than integrating the full posterior.Evaluations in LMM configurations compatible with \texttt{INLA} indicate that this approximation remains highly accurate while offering significant computational speed advantages (see Figure \ref{fig:Inla}). This efficiency facilitates the determination of the optimal hyperparameter set using the L-BFGS-B method within the R \texttt{optim} function.

Next, we outline an algorithm to approximate $\boldsymbol{\theta}^*$ the MAP estimator of the model parameters of the LMM with the priors detailed in Section \ref{subsec:LMM}. The algorithm to approximate the MAP estimator $\boldsymbol{\theta}^*$ is a variant of the Expectation-Maximization (EM) algorithm. This approach exploits the fact that the model simplifies significantly when conditioned on the latent variables $\boldsymbol{\gamma}$. At each iteration, parameters are updated via a two-step procedure:
\begin{itemize}
    \item E-step: Computes the expectation $\mathbb{E}_{\boldsymbol{\gamma}\sim p(.|\boldsymbol{\theta}_k,\mathbf{Y})} \, \{ \log[f(\mathbf{Y},\boldsymbol{\gamma} \, |\, \boldsymbol{\theta}) \, \pi(\boldsymbol{\theta} \,| \, \boldsymbol{\Theta})]\}$, relative to the current parameter values $\boldsymbol{\theta}_{k}$.
    \item M-step: Identifies the parameter update $\boldsymbol{\theta}_{k+1}$ that maximizes this expectation.
\end{itemize}The resulting updating rules are as follows:
\begin{eqnarray*}
\boldsymbol{\beta}_{k+1} & = & (\mathbf{X}^\top \mathbf{X} + \lambda \mathbf{I}_p)^{-1}\mathbf{X}^\top (\mathbf{Y}- \sum\nolimits_{r=1}^R \mathbf{Z}_r \mathbb{E}[\boldsymbol{\gamma}_{r} \, | \, \theta_{k},\mathbf{Y}] ),
\\
\sigma^2_{k+1} & = &  (p+n)^{-1} \{ \lambda \boldsymbol{\beta}_{k+1}^\top \boldsymbol{\beta}_{k+1}+ \mathbb{E}[\|\mathbf{Y}-\mathbf{X} \boldsymbol{\beta}_{k+1} - \sum\nolimits_{r=1}^R \mathbf{Z}_r \boldsymbol{\gamma}_r\|^2_2 \, | \, \theta_{k},\mathbf{Y}] \} \text{ and }
\\
\mathbf{\Sigma}_{r,k+1} & = & (m_r+\eta_r+q_r+1)^{-1} \{\boldsymbol{\Phi}_r + \sum\nolimits_{j=1}^{m_r} \mathbb{E}[\boldsymbol{\gamma}_{r,j}\boldsymbol{\gamma}_{r,j}^\top \, | \, \theta_{k},\mathbf{Y}] \}.
\end{eqnarray*}
Analytical expressions for the conditional expectations are provided in the Appendix. The algorithm iterates through these steps until convergence is achieved.

A more intuitive parameterization of the Inverse-Wishart distribution is provided to clarify the influence of hyperparameters on the MAP estimator. The update rule for $\mathbf{\Sigma}_{r,k+1}$ can be reformulated as:
\begin{eqnarray*}
\boldsymbol{\Sigma}_{r,k+1} & =&  b_r\mathbf{A}_r+(1-b_r) m_r^{-1} \sum\nolimits_{j=1}^{m_r}\mathbb{E}[\boldsymbol{\gamma}_{r,j}\boldsymbol{\gamma}_{r,j}^\top \, | \, \theta_{k},\mathbf{Y}],  
\end{eqnarray*}
where $A_r =\frac{\boldsymbol{\Phi}_r}{\eta_r+q_r+1} $ is the mode of the inverse-Wishart distribution and $b_r = \frac{\eta_r+q_r+1}{m_r+\eta_r+q_r+1}$ can be loosely viewed as the strength of the prior. Notably, the second term in this sum corresponds to the standard EM update for the maximum likelihood estimator. Consequently, $b_r = 0$ signifies an uninformative prior where the MAP aligns with the maximum likelihood estimate. Conversely, $b_r = 1$ is equivalent to a Dirac prior at the mode $\mathbf{A}_r$, rendering the posterior independent of the data.\\

\section{Application on simulated and real-life data}\label{sec:data}

This section demonstrates the efficacy of the proposed methodology through a controlled simulation study followed by an application to an empirical dataset. Consistent with the objectives outlined in the introduction, the simulation primarily investigates the impact of regularizing random-effect parameters. Additionally, we briefly illustrate the capacity of our method to perform joint regularization of both fixed and random effects.

To evaluate practical utility, the method is applied to the longitudinal dataset of \cite{van_nunen_short-term_2021}. This dataset originates from a study in which participant outcomes and time-varying covariates were measured at three distinct time points. Participants were recruited from three different cities; this shared origin introduces a hierarchical correlation structure—nesting observations within individuals and individuals within cities.

While the dataset structure may appear parsimonious, its specific constraints—namely the sparse number of repeated measurements per individual and the limited number of higher-level units (cities)—render it an ideal case study for our methodology. These characteristics typically impede stable estimation in conventional LMM frameworks.

The specific characteristics of the dataset are detailed in Section \ref{subsec:realData}. Preceding this, Section \ref{subsec:sim} utilizes simulated data generated under a similar experimental configuration to rigorously examine the empirical properties and performance of the proposed method.

\subsection{Simulation study}\label{subsec:sim}
The performance of the proposed methodology is evaluated through a simulation study, which facilitates the assessment of parameter and hyperparameter estimation quality, and predictive accuracy. This subsection focuses on the regularization of random effects within a low-dimensional fixed-effect context. Subsequent analysis demonstrates that the integration of high-dimensional fixed effects does not fundamentally alter these findings.

\subsubsection{Random effects regularisation}

In this simulation set-up, we set the number of observations to $n=240$ (roughly the same number as the study of \cite{van_nunen_short-term_2021}) from a LMM with two random effects, mimicking the individual and city-level correlations. We assume that the population comprises 60 individuals (again similar to the number of individuals in the study of \cite{van_nunen_short-term_2021}). Each individual is observed exactly four times. The 240 observations are equally balanced across four cities. Without loss of generality, but unlike the study of \cite{van_nunen_short-term_2021}, the individuals are not nested within cities (hence allowing people to move from city to city).

The simulation parameters are designed to reflect the study by \cite{van_nunen_short-term_2021}, with the number of observations set to $n=240$. The model incorporates two random effects to simulate correlations at the individual and city levels. The population comprises 60 individuals, each observed exactly four times, with observations balanced across four cities. Unlike the motivating empirical dataset, individuals in this simulation are not nested within cities, thereby allowing for simulated participant mobility between locations.

For each observation, a three-column fixed-effect design matrix $\mathbf{X}$ is constructed from an intercept and two covariates sampled from independent standard Gaussian distributions. The random-effect covariates are specified as $\mathbf{U}_1 = \mathbf{U}_2 = \mathbf{X}$, introducing random coefficients for both individual and city-level covariates. The true fixed-effect parameter $\boldsymbol{\beta}_0$ is drawn from $\mathcal{N}(0, 1)$, while both random-effect covariance matrices are initialized as $\boldsymbol{\Sigma}_{0,1} = \boldsymbol{\Sigma}_{0,2} = \boldsymbol{\Sigma}_{0}$. The residual standard deviation $\sigma_0$ is scaled to ensure that the total variance of the random effects and the error term are comparable.

Two estimators are compared: the Maximum Likelihood Estimator ($\boldsymbol{\theta}_{\text{ML}}$) and the Empirical Bayes Regularized Estimator ($\boldsymbol{\theta}_{\text{EB}}$). The EB estimator is defined as the MAP estimator, with hyperparameters $\boldsymbol{\Theta}_{\text{EB}}$ chosen to maximize the marginal likelihood. Using the reparameterization from Section \ref{subsec:LMM}, the prior mode $\mathbf{A}$ is set as a constant diagonal matrix ($a\mathbf{I}_3$). Given the low dimensionality of the fixed effects, the ridge penalty $\lambda$ is set to zero for computational efficiency. Note that, as shown in Supplementary Figure \ref{fig:highLow}, when fixed effects are low-dimensional, joint regularization correctly yields a negligible estimate for $\lambda$.

\begin{figure}[h!]
    \centering
    \includegraphics[scale = 0.23]{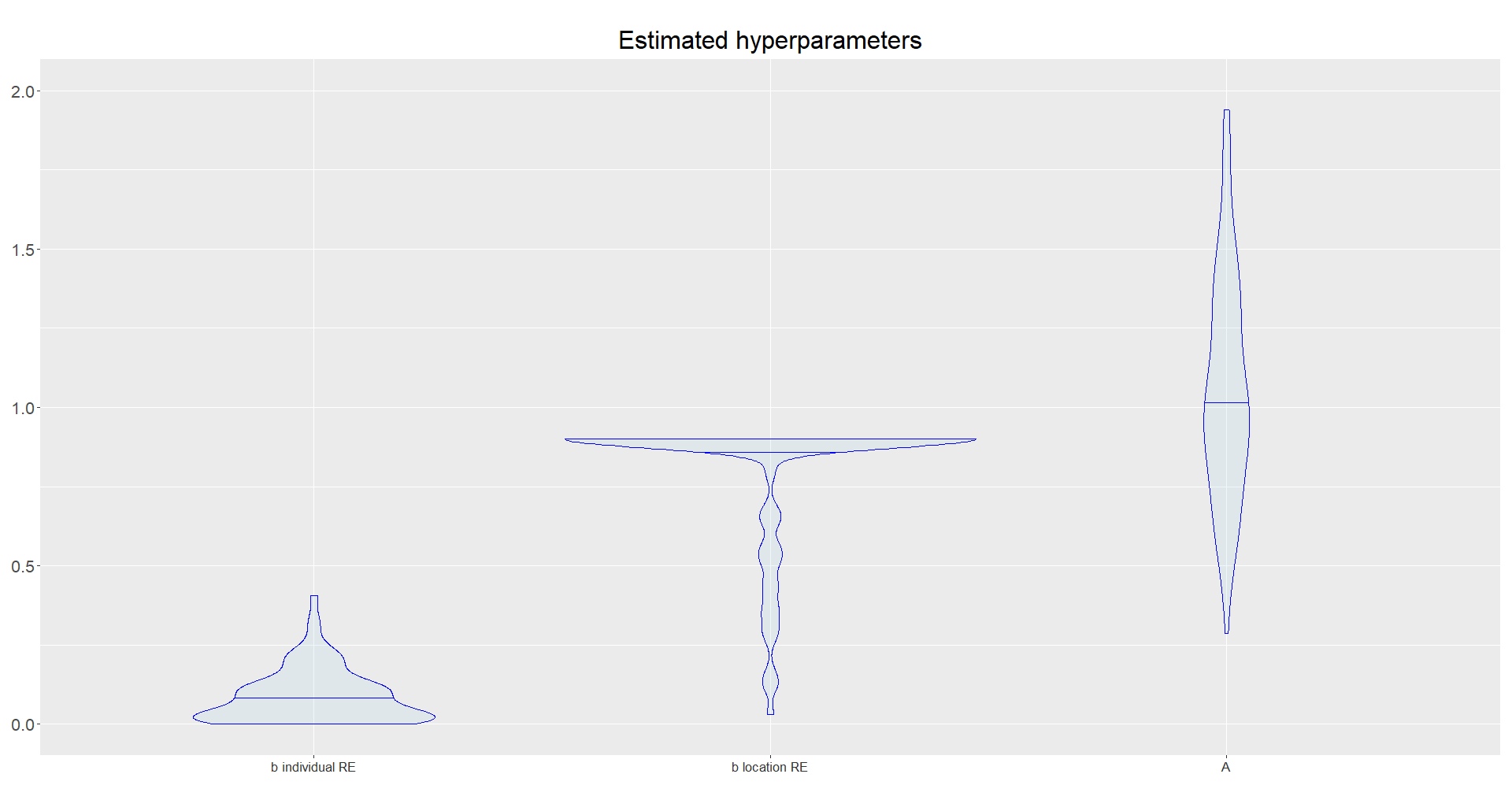}
    \caption{Violinplot of the estimated hyperparameters of the IW-prior of the covariance of the random effects over 150 simulated data.}
    \label{fig:hp}
\end{figure}

Prior to comparing the estimators, the convergence of the EB hyperparameters was assessed. Figure \ref{fig:hp} illustrates the distribution of the learned hyperparameters $\boldsymbol{\Theta}_{\text{EB}} = (b_1, b_2, a)$ across the 150 experiments. The estimated scale parameter $a$ is concentrated near unity. Given that the identity matrix serves as the closest approximation to the true covariance $\boldsymbol{\Sigma}_0$, this centering indicates that the estimator effectively shrinks toward the optimal covariance structure. Regarding the prior strength parameters $b_1$ and $b_2$:\begin{itemize}
    \item Individual Level ($b_1$): Estimates remain consistently low, indicating that the individual random-effect estimation relies primarily on the large number of subjects in the dataset rather than the prior.
    \item City Level ($b_2$): Estimates approach the numerical upper bound of 0.9. This high degree of regularization is a response to the limited number of levels (cities), which necessitates prior-driven stability.
\end{itemize}

The potential interdependency of the hyperparameter estimates was investigated, with comprehensive details provided in the Supplementary Material and summarized herein. To determine whether the $b$ parameters exert mutual influence, the simulation was replicated using a model specification containing only a single random effect. This comparative analysis confirms that the concurrent inclusion of multiple random effects does not perturb the hyperparameter estimates of the individual model components

\begin{figure}[h!]
    \centering
    \includegraphics[scale = 0.23]{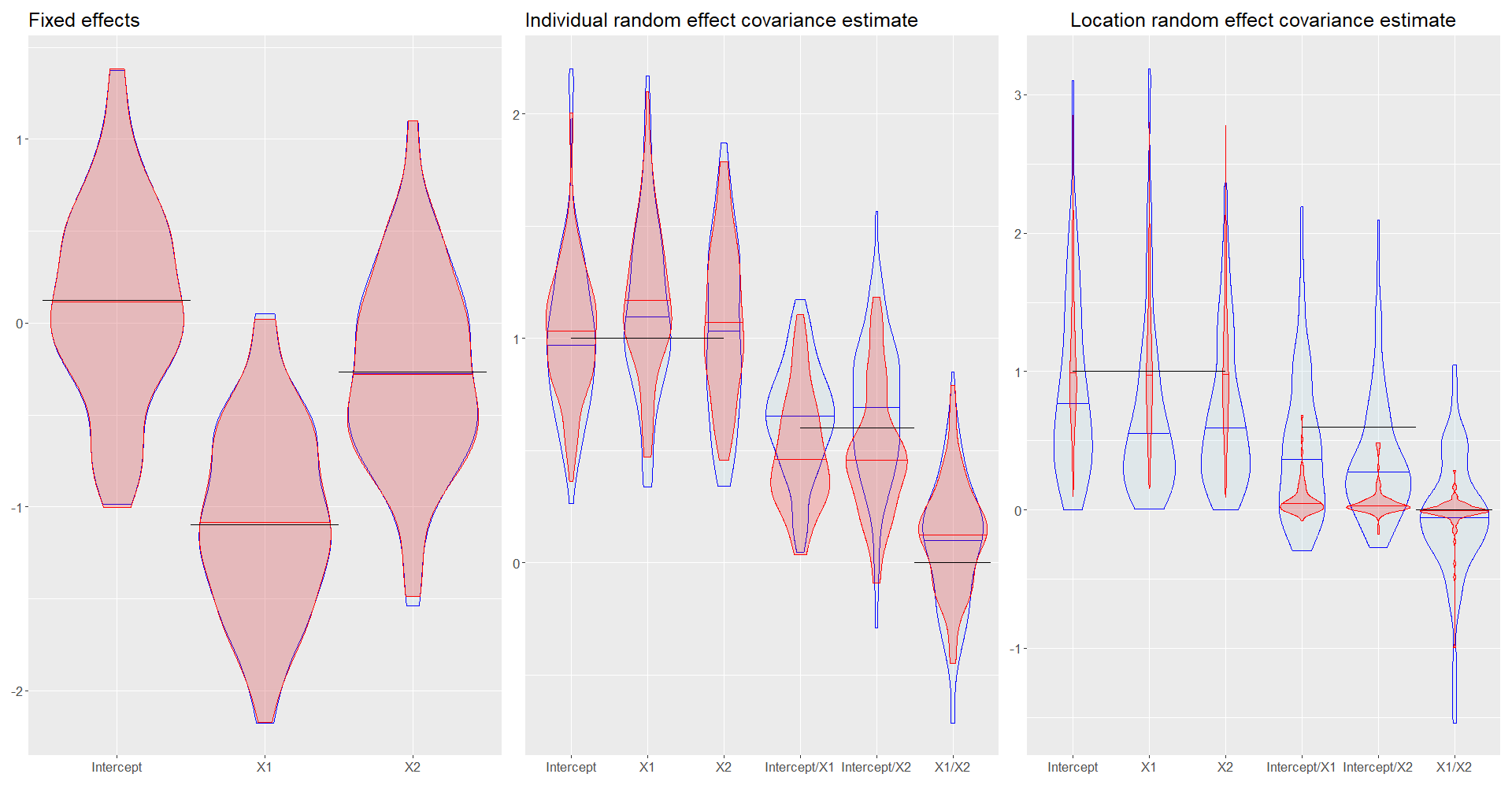}
    \caption{Violinplot of the MAP (in red) and maximum likelihood (in blue) estimates of the fixed and random effects over 150 simulated data. Black lines show the value of the true parameter. }
    \label{fig:simParam}
\end{figure}

We now direct our attention towards the comparison of the parameter estimates $\hat{\boldsymbol{\theta}}_{\text{ML}}$ and $\hat{\boldsymbol{\theta}}_{\text{EB}}$, first for the random effects parameters. The panels of Figure \ref{fig:simParam} show the violinplots of the two estimates and the true parameter values. The regularized estimator is biased, although not dramatically, but it exhibits substantially less variance than the (almost unbiased) maximum likelihood estimator. This bias-variance trade-off, is particularly evident in the case of off-diagonal terms (see Figure \ref{fig:simParam}). This is due to the influence of the hyperparameters $b$ on the maximum a posteriori estimate. As $b_1$ is small for the individual random effect, there is minimal disparity between the maximum likelihood and MAP estimators. Conversely, for the city random effect, where regularization $b_2$ is substantially stronger, the MAP estimator for the covariance of the random effect is predominantly set to $A$, resulting in less variance.

The two estimates of the fixed effect $\boldsymbol{\beta}$ show little difference (Figure \ref{fig:simParam}), irrespective of the differences in the random effect variance matrices between the two estimators. This robustness of the fixed effect estimates to that of the random effect has previously been reported. \cite{verbeke_effect_1997}, demonstrated that misspecification of the random effect distribution has minimal consequences on fixed effect estimation. Similarly, \cite{taylor_does_1998}, concluded that the covariance structure has no bearing on the estimation of fixed effect parameters. 

\begin{figure}[h!]
    \centering
    \includegraphics[scale = 0.22]{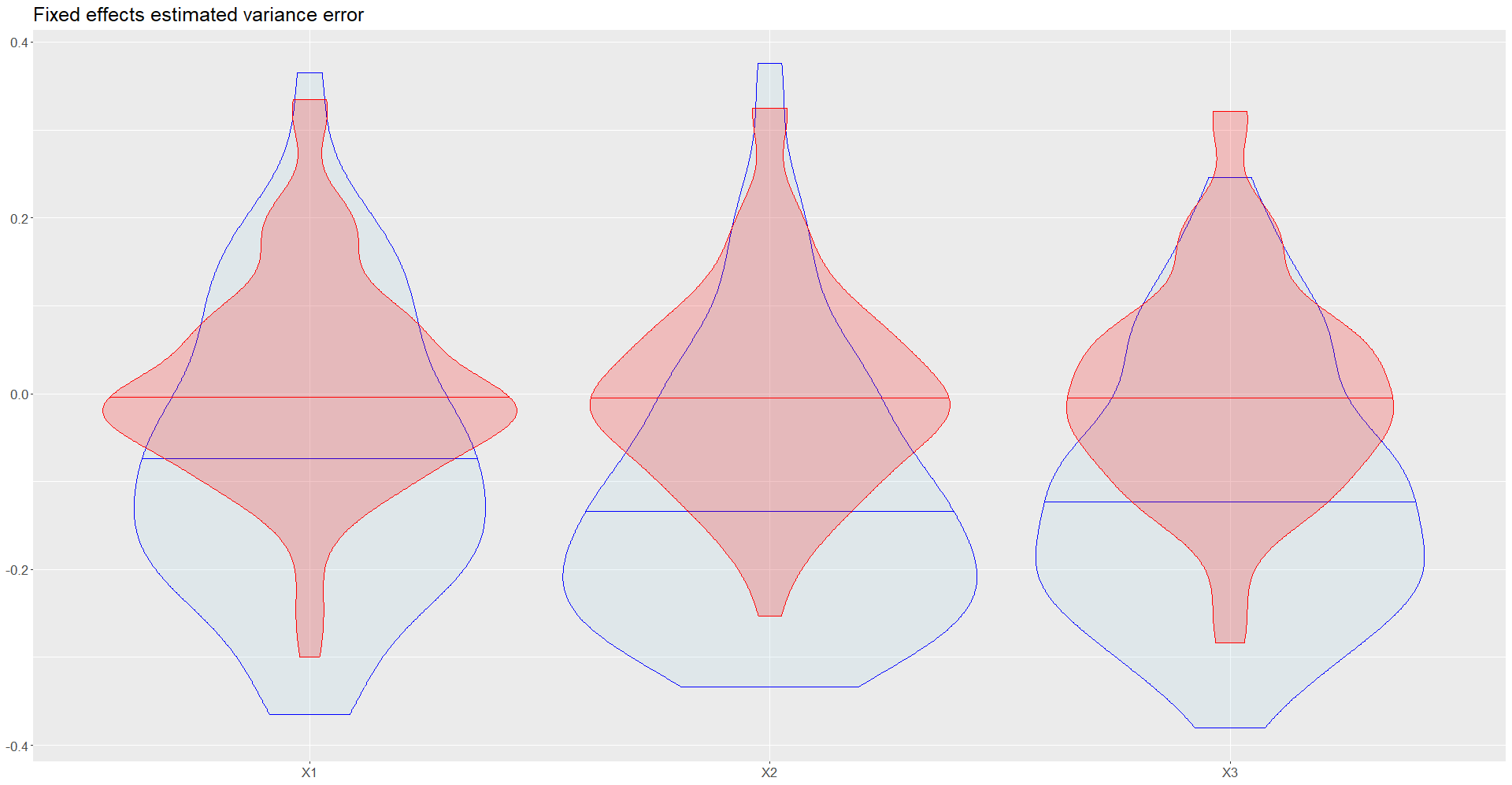}
    \caption{Violinplot of the estimation error of the MAP (in red) and maximum likelihood (in blue) fixed effects variance associated with the covariates ($X_1,X_2,X_3$) with over 150 simulated data. }
    \label{fig:stdBeta}
\end{figure}

While fixed-effect point estimates are robust, the accuracy of the covariance estimates is critical for quantifying uncertainty. To perform hypothesis testing or calculate $p$-values (e.g., via the lmerTest package), one must accurately estimate the variance of the fixed-effect estimator, defined as:$$c(\boldsymbol{\theta}) = \left[ \mathbf{X}^\top \left( \sigma^2 \mathbf{I}_n + \sum_{r=1}^{R} \mathbf{Z}_r \mathbf{W}_r \mathbf{Z}_r^\top \right)^{-1} \mathbf{X} \right]^{-1}.$$Figure \ref{fig:stdBeta} presents the estimation error $c(\boldsymbol{\theta}_0) - c(\hat{\boldsymbol{\theta}})$. The EB-regularized estimator demonstrates a smaller error than the ML estimator and remains essentially unbiased. This indicates that incorrect covariance estimation in non-regularized models directly degrades the reliability of subsequent statistical inference and significance testing.

Lastly, we compare the predictive performance of our two estimates. We focus on the distribution of $Y_{new}|Y_{obs}$, where $Y_{new}$ represents the unknown outcome associated with a known set of design matrices $D_{new}$ and $Y_{obs}$ denotes the observed outcomes associated with the design matrices $D_{obs}$. The conditional distribution  $Y_{new}|Y_{obs}$ can be derived in closed form using the estimated parameters, as $Y_{new}$ and $Y_{obs}$ follow a joint normal distribution. Its evaluation only requires $X_{obs}$, $Y_{obs}$ and $X_{new}$. We choose these in two ways, the first choice i) facilitates the assessment of the model's ability to predict new measurements of individuals and cities already present in the dataset, whereas the second ii) does so for individuals and cities that are not part of the original data.
\begin{compactitem}
\item[i)] Let $Y_{obs}$ and $X_{obs}$ be the data used for parameter estimation. The $X_{new}$ is generated in the same way as $X_{obs}$.

\item[ii)] Generate $Y_{obs}$ and $X_{obs}$ independent of the training data. The $X_{new}$ is generated in the same way as $X_{obs}$. The design matrices of the second observed dataset are comparable in terms of their distribution to the one in the training dataset. The case where the design matrices differ can be found in the appendix.
\end{compactitem}
We evaluate the predictive power using the Kullback-Leibler (KL) divergence between the estimated distribution $p_{\hat{\boldsymbol{\theta}}}( Y_{\text{new}}| \mathbf{Y_{\text{obs}}})$ and the true distribution $p_{\boldsymbol{\theta}_0}( Y_{\text{new}}| \mathbf{Y_{\text{obs}}})$. We use the KL divergence as it combines the evaluation of the point estimate for the new observations and the variance surrounding the prediction. The root mean squared error (RMSE) of the conditional point prediction, defined as $\text{RMSE}(\hat{\boldsymbol{\theta}}) = ||\mathbb{E}_{\hat{\boldsymbol{\theta}}}(\mathbf{Y_{\text{new}}}| \mathbf{Y_{\text{obs}}}) - \mathbb{E}_{\boldsymbol{\theta}_0} (\mathbf{Y_{\text{new}}}| \mathbf{Y_{\text{obs}}}  )||^2$ for each scenario can be found in the Supplementary material.  
Figure \ref{fig:KLRMSE} presents a violin plot illustrating the difference in scores between the non-regularized and regularized estimators for the 150 experiments, while keeping $D_{obs}$ and $D_{new}$ fixed. A higher positive value indicates better performance by our regularized estimator when predicting the outcomes.

\begin{figure}
    \centering
    \includegraphics[scale = 0.2]{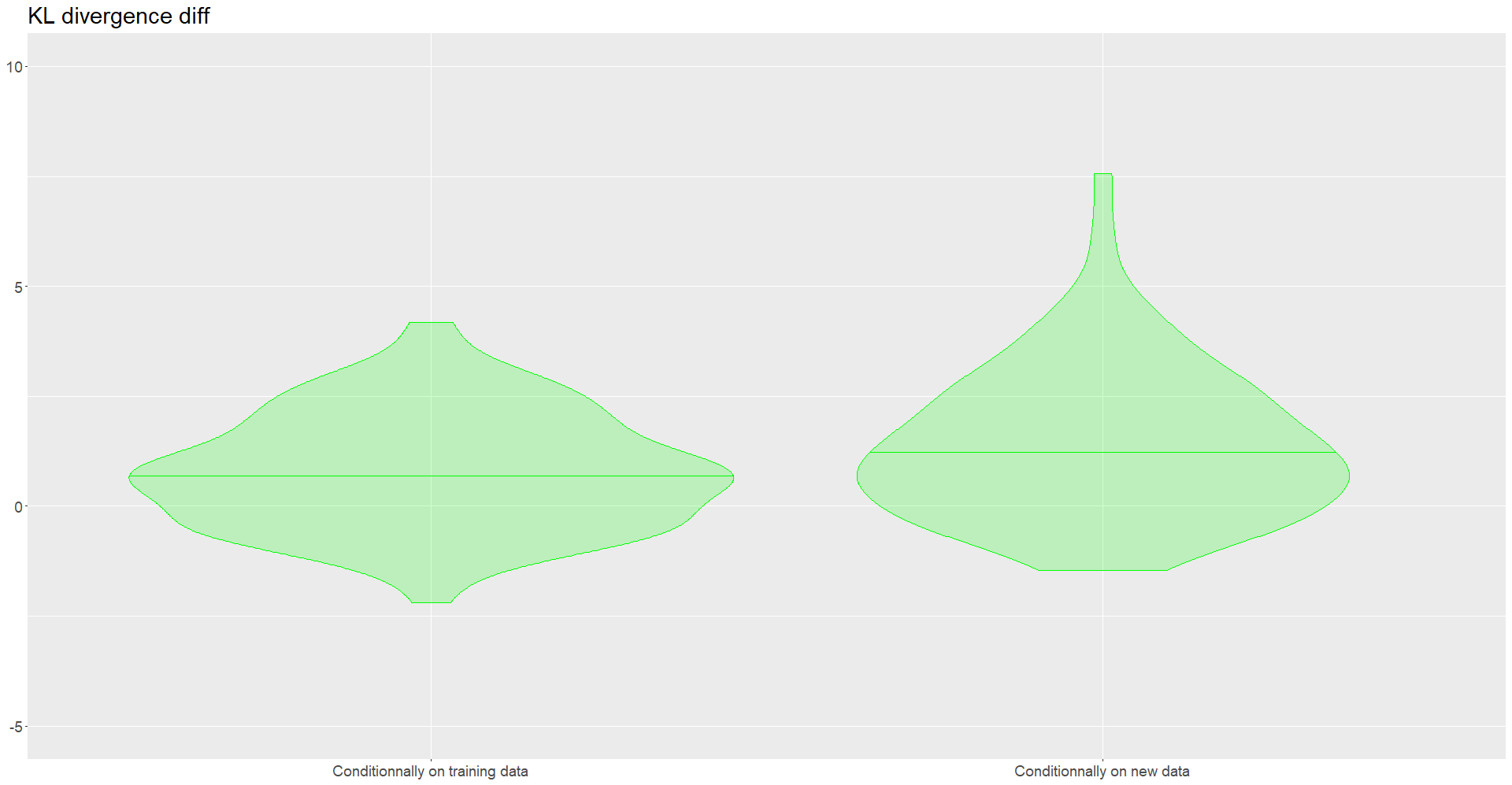}
    \caption{Violinplot of the KL divergence difference $\text{KL}[p_{\hat{\boldsymbol{\theta}}_{\text{ML}}}( Y_{\text{new}}| \mathbf{Y_{\text{obs}}}), p_{\boldsymbol{\theta}_0}( Y_{\text{new}}| \mathbf{Y_{\text{obs}}})]-\text{KL}[p_{\hat{\boldsymbol{\theta}}_{\text{EB}}}( Y_{\text{new}}| \mathbf{Y_{\text{obs}}}), p_{\boldsymbol{\theta}_0}( Y_{\text{new}}| \mathbf{Y_{\text{obs}}})]$ for both scenarios over the 150 experiments. Positive values indicate superior performance of the joint regularization approach. }
    \label{fig:KLRMSE}
\end{figure}

The regularized estimator consistently outperforms the non-regularized estimator over all the examined scenarios and scores. Note that the scenarios exhibit substantial performance differences. Intuitively, this can be understood when realizing that the random effects' design matrices employed in the estimation and prediction are the same in the first scenario, while they differ in the second one. This benefits the regularized estimators as shrinkage methods usually generalize better to new cases. 
% This improved performance exhibits itself in both metrics, only in the first scenario both estimators have a comparable RMSE. Overall the performance improvement is more pronounced in the KL divergence metric. This disparity between the RMSE and KL divergence can be explained by their construction. The RMSE solely relies on the predicted conditional mean, where the estimates of the random effect covariance play a minor role. On the other hand, the KL divergence incorporates the complete conditional predictive distribution, giving more weight to the variance of the random effects.

\subsubsection{Joint regularisation with high dimensional fixed effexts}
This section demonstrates the capacity of the proposed method for the joint regularization of both fixed and random effects. In contrast to the preceding analysis, we consider a configuration involving 300 fixed-effect covariates. To mitigate computational demands, the analysis is restricted to a single random effect. While the random effect modeling in this scenario is simpler than in the preceding one, it maintains a reasonable level of complexity by incorporating a random intercept and two random slopes. Notably, the covariates defining the random-slopes design matrix, $\mathbf{U}$, are distinct here from those utilized in the fixed-effects component, $\mathbf{X}$.

\begin{figure}[h!]
    \centering
    \includegraphics[width=0.8\textwidth]{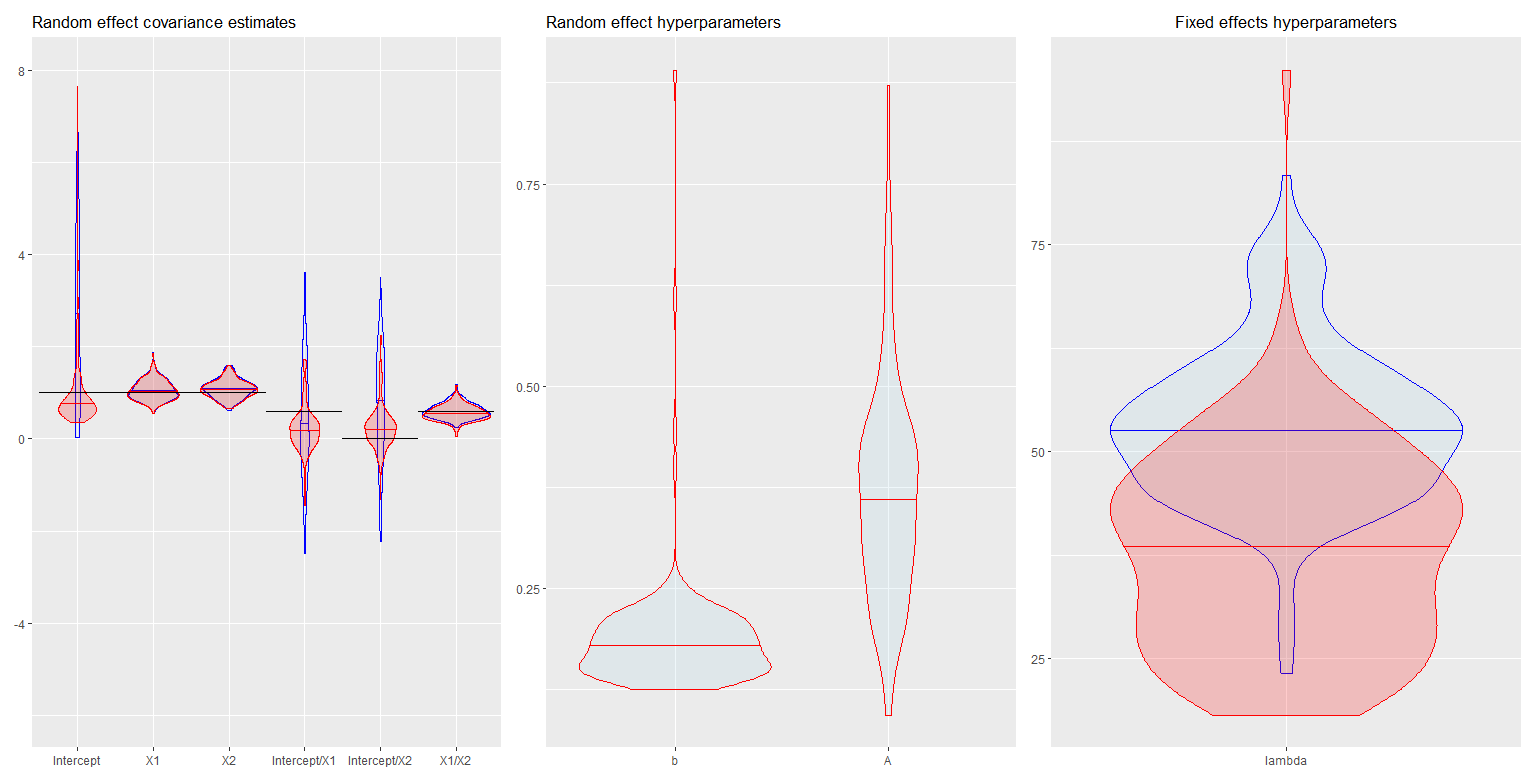}
    \caption{Violin plots of estimated random-effect parameters and hyperparameters under high-dimensional fixed effects ($p=300, n=240$). The blue plots represent regularization of fixed effects only (with the random-effect strength parameter $b$ set to $0$), while the red plots represent joint optimization of hyperparameters for both fixed and random effects. The black lines denote the ground truth parameters.}
    \label{fig:highDima}
\end{figure}

Given the high dimensionality of the fixed effects, a direct comparison between the EB and ML estimation is no longer feasible. Consequently, we define two regularization scenarios: one in which only the fixed-effect parameters are regularized, and a second involving the joint regularization of both fixed and random effects. Figure \ref{fig:highDima} illustrates the learned hyperparameters for both scenarios alongside the resulting random-effect covariance matrix estimates. We observe that the regularization parameter $\lambda$ operates on a comparable scale across both settings. Furthermore, the estimated hyperparameters for the random effects, as well as the corresponding covariance estimates, exhibit behavior consistent with the results observed in the previous section. 

\begin{figure}[h!]
    \centering
    \includegraphics[scale = 0.23]{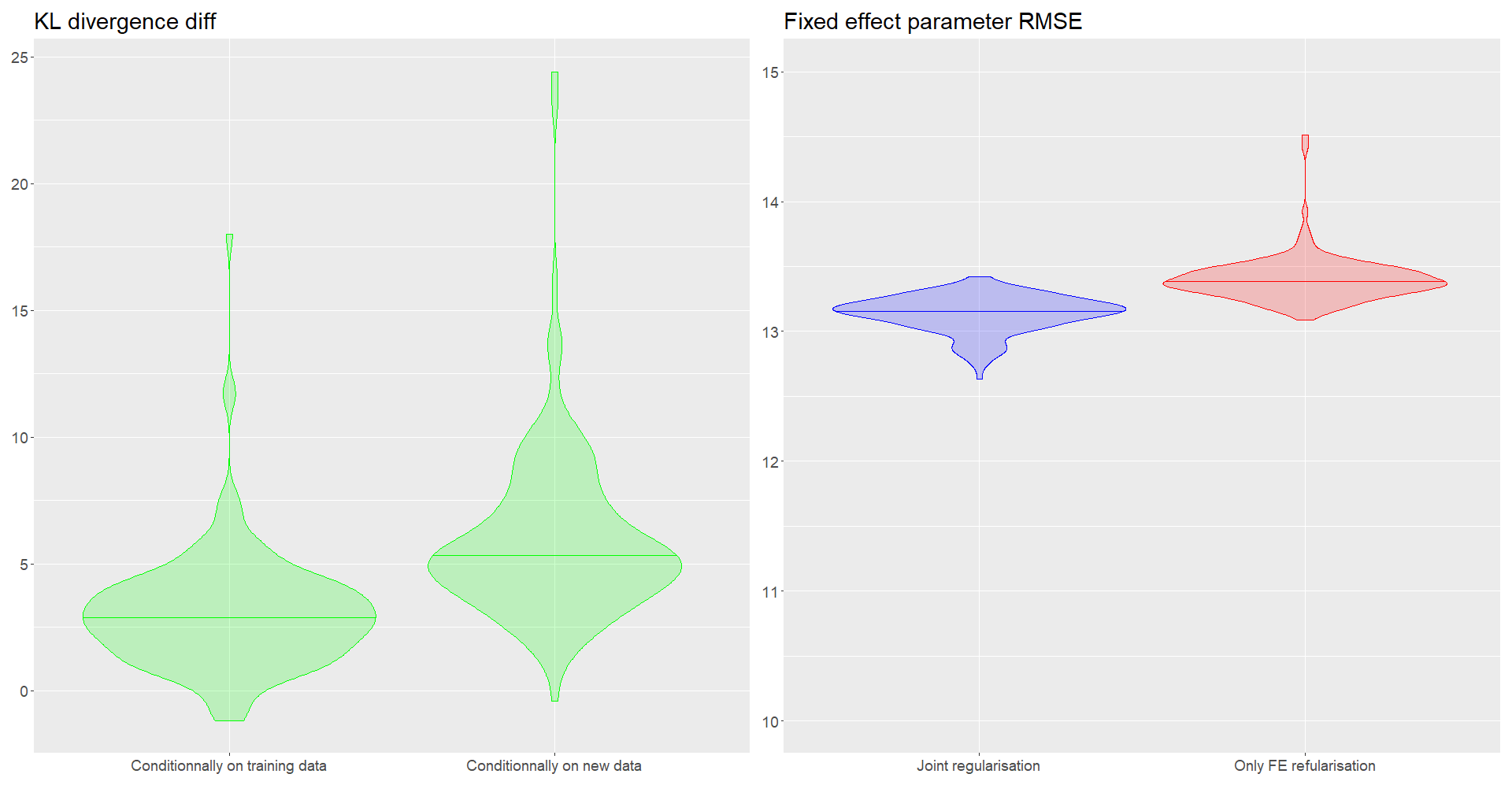}
    \caption{Right: Violin plots of the RMSE for the fixed-effect estimates. Estimates under fixed-effect regularization only are shown in red, while joint regularization results are shown in blue. Left: Violin plots of the difference in KL divergence between the joint regularization and fixed-effect-only scenarios. Positive values indicate superior performance of the joint regularization approach.}
    \label{fig:hdKLbeta}
\end{figure}

We now evaluate the performance of the estimators. Figure \ref{fig:hdKLbeta} presents the distribution of the RMSE for the fixed-effect estimates (right panel) and the gain in KL divergence for prediction when employing joint regularization relative to fixed-effect regularization alone (left panel). We observe that the error in fixed-effect parameters is similar across both scenarios. This result is expected, as the hyperparameter $\lambda$ remains relatively stable in both cases and, as previously established, the regularization of random effects exerts limited influence on fixed-effect estimation. Crucially, however, the predictive benefits caused by regularizing the random effects are consistently observed. 

These findings corroborate the results from the previous section, confirming the advantages of regularizing the random-effect covariance matrix. Furthermore, they demonstrate the efficacy of our proposed procedure in handling the simultaneous regularization of both parameter sets.

\subsection{Application evaluating health impact of air pollution}\label{subsec:realData}
We apply our methodology to the dataset presented in \cite{van_nunen_short-term_2021}, which studies the connection between health outcomes and ultra-fine and fine air pollution particles. The dataset comprises longitudinal observations involving 107 individuals distributed across three locations (the Netherlands, Basel, and Turin) on three different days spanning a year. Measurements of air pollutants fine particulate matter (PM25), soot, and ultrafine particles (UFP) are obtained through various sensors. Blood pressure and lung function assessments were conducted at the end of each observation day. Although the study investigates associations between various pollutant combinations and health effects, we focus on the link between soot exposure and diastolic blood pressure. This prioritization is based on the paper's identification of this relationship as statistically significant.

The dataset's structure allows two potential layers of correlation among observations: at the location level and at the individual level. The standard approach for modeling such correlated observations involves an LMM with two random effects: one for the location level and another for the individual level. However, the limited number of observed levels at the city level (only three) and the scarcity of data pose challenges to accurately estimate such a complex model. 
% \textit{In the original analysis of \cite{van_nunen_short-term_2021} these challenges are overcome by conducting separate regressions for each location and combining the location-specific estimates through a meta-analysis technique to estimate the overall effect.} 
In the original analysis of \cite{van_nunen_short-term_2021}, these challenges are overcome by estimating, separately for each location, an LMM with a random intercept at the individual level, and combining these estimates with a meta-analysis to estimate the overall effect. Such an analysis has two limitations. Firstly, the division of data and subsequent aggregation through meta-analysis is artificial as a single comprehensive analysis is feasible. Secondly, the variance estimation of fixed effect $\boldsymbol{\beta}$ in a `per-location' regression is inconsistent. Our simulation study demonstrates that, if the variance structure of the random effects is inadequately specified or estimated, the variance estimation of the fixed effect $\boldsymbol{\beta}$ is affected. Consequently, this hinders the proper aggregation of results during the meta-analysis step. These limitations motivate us to compare the meta-analysis methodology with our framework, which incorporates two random effects (i.e. city and individual).

We compare three models in our analysis. The first model, $Y\sim X+(1 \, | \, \text{ind})$, is the one utilized in the paper \cite{van_nunen_short-term_2021}. It is fitted independently for each city, and the fixed effect coefficient for exposure is pooled using a fixed effect meta-analysis. This resulting estimator is referred to as $\boldsymbol{\theta}_{\text{base}}$ and serves as the baseline estimator. The second model aims to closely resemble the baseline model while avoiding the meta-analysis step. It includes an additional random effect for location: $Y\sim X+(1 \, | \, \text{ind})+(1 \, | \, \text{loc})$. We fit this model both with and without regularization, resulting in the estimators $\boldsymbol{\theta}_{\text{simple}}$ and $\boldsymbol{\theta}_{\text{simple}}^{\text{reg}}$, respectively. Finally, to explore the benefits of regularization in fitting more complex models, we incorporate temperature as a covariate in the individual-level random effect and fit the model $ Y\sim X+(1+\text{temp} \, | \, \text{ind})+(1 \, | \, \text{loc}) $ with and without regularization, leading to the estimators $\boldsymbol{\theta}_{\text{complex}}$ and $\boldsymbol{\theta}_{\text{complex}}^{\text{reg}}$.

We evaluate the aforementioned estimators by cross-validation, partitioning the data set into a training set and a test set. We randomly select one observation per individual to compose the test data set $\mathcal{D}_{\text{test}}$, while the remaining observations form the training set $\mathcal{D}_{\text{train}}$. We then estimate the parameters and hyperparameters (for the regularized models only), using the training set and evaluate the performance of the estimation on the test set. To assess the quality of the estimates $\boldsymbol{\theta}_{\text{base}},\boldsymbol{\theta}_{\text{simple}},\boldsymbol{\theta}_{\text{simple}}^{\text{reg}},\boldsymbol{\theta}_{\text{complex}},\boldsymbol{\theta}_{\text{complex}}^{\text{reg}}$, we compute the RMSE of the predicted outcomes on the test set, conditioned on the training set: $RMSE(\boldsymbol{\theta}) =\| \widehat{\mathbf{Y}}_{\boldsymbol{\theta}}-\mathbf{Y}_{\text{test}}\|^2/n_{\text{test}}$, where $\widehat{\mathbf{Y}}_{\boldsymbol{\theta}}=\mathbb{E}[\mathbf{Y}|\mathcal{D}_{\text{train}},\mathbf{X}_{\text{test}},\boldsymbol{\theta}]$. \\

\begin{figure}[ht!]
    \centering
    \includegraphics[scale = 0.35]{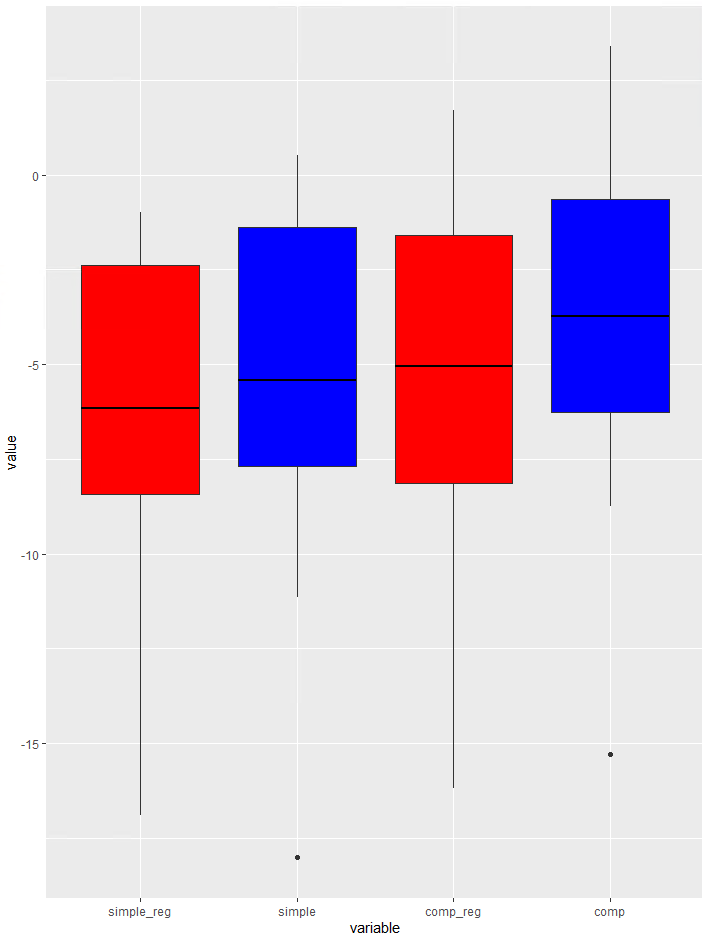}
    \caption{Boxplot of the RMSE of the prediction for the regularised estimators (red) and their non-regularized counterparts (red) for the complex and simple models over 10 random train/test splits. The RMSE is relative to the one of the baseline model described in \cite{van_nunen_short-term_2021}. }
    \label{fig:true}
\end{figure}
We assess the stability of the RMSE through multiple experiments involving different random train-test splits. Figure \ref{fig:true} presents the relative (to the baseline model described in the reference paper) improvement in RMSE for each of the four models: simple and complex models, both in non-regularized form. All models exhibit lower RMSE values than the baseline. This substantiates the utilization of the entire dataset for estimation, instead of aggregating estimates via a meta-analysis approach, for a more robust predictive model.

Our stability analysis indicates that the regularized estimators outperform their non-regularized counterparts. The magnitude of improvement due to regularization is more pronounced in the context of more complex models. Specifically, the complex regularized model achieves a performance level equivalent to that of the non-regularized simple model. This observation underscores the capacity of our empirical Bayesian procedure to estimate more complex models without sacrificing predictive efficacy.

\section{Discussion}
This research introduces a versatile, data-driven framework for the concurrent regularization of fixed and random effects within a LMM structure. The methodology employs empirical Bayes estimation, with hyperparameters derived through the optimization of the marginal likelihood. The development of an approximate marginal likelihood computation using a Laplace approximation renders the method practically applicable in in large dimensions. Furthermore, the intuitive reparameterization of the inverse-Wishart prior offers a robust mechanism for constructing complex random-effect covariance structures.

Simulation results indicate that this methodology significantly improves both the prediction of novel observations and the estimation of fixed-effect variance. These outcomes were validated through application to a real-world dataset, showcasing that regularization enhances predictive performance. The results suggest that regularization facilitates the fitting of intricate models without compromising statistical accuracy. Moreover, the hyperparameter estimates produced by our empirical Bayes method can be used in a full Bayesian estimation framework, next to calculation of the MAP estimator as is done in this work. This can straightforwardly be done within packages like INLA or RSTAN, enabling the derivation of the complete posterior distribution for the problem.

Two prospective extensions of this work are identified. Primarily, the method may be adapted to handle multiple independent outcomes or datasets that share common priors, leveraging the additive nature of the marginal likelihood. Additionally, the approach can be modified to encompass Generalized Linear Mixed Models (GLMMs) and the utilization of non-conjugate priors. These advancements will require the development of efficient MAP estimation procedures and suitable adaptations of the Laplace approximation to ensure computational feasibility across diverse statistical settings.

\clearpage
\printbibliography
\clearpage
\section{Appendix}

\subsection{Mathematical derivations}
In this section we first give the details of the EM algorithm to find the MAP and its adaptation to compute the maximum likelihood. We then derive the Hessian of the Log likelihood of an LMM. 

We recall that an LMM is defined by the following equation:  
\begin{align*}
\mathbf{Y}  =  \mathbf{X}\betab+ \mathbf{Z} \boldsymbol{\gamma} +\boldsymbol{\varepsilon}
\, \, \, & = \, \, \,  \mathbf{X}\betab+\sum\nolimits_{r=1}^R \mathbf{Z}_r \gammab_r +\boldsymbol{\varepsilon}\\
&  = \, \, \,  \mathbf{X}\betab+\sum\nolimits_{r=1}^R \sum\nolimits_{j=1}^{m_r} (\mathbf{K}_{r,j}\bullet\mathbf{U}_r) \gammab_{r,j} +\boldsymbol{\varepsilon},
\end{align*}
where $\boldsymbol{\gamma} = (\boldsymbol{\gamma}_1^{\top}, \ldots, \boldsymbol{\gamma}_R^{\top})^{\top}$ with the $r$-th random effect $\gammab_r = (\gammab_{r,1}^\top,\cdots,\gammab_{r,m_r}^\top)^\top \in \mathbb{R}^{m_r q_r}$.  The random effects are independent and follow $\gammab_{r,i}\sim\No(\boldsymbol{0}_{q_r},\boldsymbol{\Sigma}_r)$. We denote the variance of $\gammab_r$ by $\mathbf{W}_r $ composed of $m_r$ repeated blocks of $\boldsymbol{\Sigma}_r$, i.e. $\mathbf{W}_r=\mathbf{I}_{m_r}\otimes \boldsymbol{\Sigma}_r$, and $\boldsymbol{\varepsilon}\sim \No(\boldsymbol{0}_n,\sigma^2 \mathbf{I}_{n})$. In the following we will denote $\mathbf{K}_{r,j}\bullet\mathbf{U}_r = \mathbf{U}_{r,j}$. Additionally we set the priors to: 
\begin{eqnarray}
\boldsymbol{\beta} \sim  \mathcal{N}(0_p, \sigma^2 \lambda^{-1} \mathbf{I}_p),\,\,
    \mathbf{\Sigma}_r \sim   \mathcal{W}^{-1}(\eta_r,\boldsymbol{\Phi}_r),\,\,
    \sigma^2 \sim  \mathcal{IG}(\alpha,\alpha).
\end{eqnarray}
We denote $\boldsymbol{\theta} = (\boldsymbol{\beta},\mathbf{\Sigma_1},\cdots,\mathbf{\Sigma_R},\sigma^2)$ the parameters of the model and $\boldsymbol{\Theta} = (\lambda,\eta_1,\cdots,\eta_R,\Phi_1,\cdots,\Phi_R,\alpha)$ the parameters of the prior.

\subsubsection{Maximisation of an LMM posterior with an expectation maximisation algorithm}

There is no closed formed solution for the maximum a posteriori and we therefore resort to an EM algorithm compute it.  The EM algorithm alternates until convergence between two steps, namely:
\begin{itemize}
\item \textbf{E-step} compute $f_{\boldsymbol{\theta}^{(k)}}(\boldsymbol{\theta}) = \, \mathbb{E}_{\gammab \sim p(.|\boldsymbol{\theta}^{(k)},\mathbf{Y})} \big\{ \log \big[ p(\mathbf{Y},\gammab \, | \, \boldsymbol{\theta}) \, \pi(\boldsymbol{\theta} \, | \, \boldsymbol{\Theta}) \big] \big\}$,
\item \textbf{M-step} update \,   $\boldsymbol{\theta}^{(k+1)} =\, \argmax_{\boldsymbol{\theta}} f_{\boldsymbol{\theta}^{(k)}}(\boldsymbol{\theta})$.
\end{itemize}
Below we detail these two steps.
\subsubsection*{E-step}
We have  $\log \big[ p(\mathbf{Y},\gammab \, | \, \boldsymbol{\theta}) \, \pi(\boldsymbol{\theta} \, | \, \boldsymbol{\Theta}) \big] = \log \big[ p(\mathbf{Y} \, | \, \boldsymbol{\theta}, \gammab) \big]+\log\big[\pi(\gammab \, | \, \boldsymbol{\theta})\big] + \log\big[p(\boldsymbol{\theta}\, |\, \boldsymbol{\Theta})\big]$. We detail each summand using the fact that $\mathbf{Y} \, | \, \theta,\boldsymbol{\gamma}\sim\mathcal{N}(\mathbf{X}\boldsymbol{\beta} + \sum_{r=1}^R\sum_{j=1}^{m_r} \mathbf{U}_{r,j} \gammab_{r,j},\sigma^2I_n)$, $\gammab_{r,i}\sim\No(\boldsymbol{0}_{q_r},\boldsymbol{\Sigma}_r)$, which yields:
\begin{eqnarray*}
\log[ p(\mathbf{Y} \, | \, \boldsymbol{\theta},\boldsymbol{\gamma})] & = & c -\frac{n}{2}\log(\sigma^2)-\frac{1}{2\sigma^2}\|\mathbf{Y}-\mathbf{X}\boldsymbol{\beta} - \sum_{r=1}^R\sum_{j=1}^{m_r} \mathbf{U}_{r,j} \gammab_{r,j} \|^2_2
\\
& =& c -\frac{n}{2}\log(\sigma^2)-\frac{1}{2\sigma^2}\big[\|\mathbf{Y}-\mathbf{X}\boldsymbol{\beta} \|^2_2
- 2 \big(\mathbf{Y}-\mathbf{X}\boldsymbol{\beta}\big)^\top \sum_{r=1}^R\sum_{j=1}^{m_r} \mathbf{U}_{r,j} \gammab_{r,j}
\\
& & +\sum_{r,r'=1}^R\sum_{j,j'=1}^{m_r}\Tr( \mathbf{U}_{r',j'}^\top \mathbf{U}_{r,j}  \boldsymbol{\gamma}_{r,j} \boldsymbol{\gamma}_{r',j'} ^\top)\big],
\\
% =&c -\frac{n}{2}\log(\sigma^2)-\frac{1}{2\sigma^2}\big[\|\mathbf{Y}-\mathbf{X}\boldsymbol{\beta} \|^2_2
% -2\big(\mathbf{Y}-\mathbf{X}\boldsymbol{\beta}\big)^\top \sum_{r=1}^R\sum_{j=1}^{m_r} \mathbf{U}_{r,j} \gammab_{r,j}\big]\\
\log[ p(\boldsymbol{\gamma} \, | \, \boldsymbol{\theta})] & =& c-\frac{1}{2} \sum_{r=1}^R\sum_{j=1}^{m_r}\left(\log(|\boldsymbol{\Sigma}_r|)+\boldsymbol{\gamma}_{r,j}^\top \boldsymbol{\Sigma}_r^{-1}\boldsymbol{\gamma}_{r,j}\right)
\\
&=&c -\frac{1}{2} \sum_{r=1}^R\left(m_r\log(|\boldsymbol{\Sigma}_r|)+\Tr(\boldsymbol{\Sigma}_r^{-1}\sum_{j=1}^{m_r}\boldsymbol{\gamma}_{r,j}\boldsymbol{\gamma}_{r,j}^\top)\right),
\\
\log \big[ \pi(\boldsymbol{\theta} \, | \, \boldsymbol{\Theta})\big] & =& c-\frac{p}{2}\log(\sigma^2)-\frac{\lambda}{2\sigma^2}\boldsymbol{\beta}^\top\boldsymbol{\beta}-(\alpha+1)\log(\sigma^2) -\frac{\alpha}{\sigma^2} 
\\
&  & -\sum_{r=1}^R\frac{\eta_r+q_r+1}{2}\log(|\boldsymbol{\Sigma}_r|)-\frac{1}{2} \Tr( \boldsymbol{\Sigma}_r^{-1}\boldsymbol{\Phi}_r).
\end{eqnarray*}
In the above we have grouped all the terms that do not depend on $\boldsymbol{\theta}$ in a constant $c$ that can differ from one equality to the other. 

To compute $f_{\theta^{(k)}}(\boldsymbol{\theta}) = \mathbb{E}_{\gammab\sim p(.|\boldsymbol{\theta}^{(k)},\mathbf{Y})} \big\{ \log \big[ p(\mathbf{Y},\gammab \, | \, \boldsymbol{\theta}) \, \pi(\boldsymbol{\theta} \, | \, \boldsymbol{\Theta})\big] \big\}$ we need an expression for $\boldsymbol{\mu}_{r,j}^{(k)} = \mathbb{E}_{\gammab\sim p( \cdot \, | \, \boldsymbol{\theta}^{(k)},\mathbf{Y})} ( \gammab_{r,j} )$ and  $\boldsymbol{\Omega}_{r,j}^{(k)} = \mathbb{E}_{\gammab\sim p(.|\boldsymbol{\theta}^{(k)},\mathbf{Y})} ( \gammab_{r,j}\gammab_{r,j}^\top )$. These two quantity can be derived noting that the $\gammab_{r,j}$ and $\mathbf{Y}$ are jointly normal:
\begin{eqnarray*}
\begin{bmatrix} \gammab_{r,j}\\
\mathbf{Y}
\end{bmatrix} | \, \boldsymbol{\theta}^{(k)} & \sim & \mathcal{N}\left(
\begin{bmatrix}
0\\
\mathbf{X}\boldsymbol{\beta}^{(k)}
\end{bmatrix},
\begin{bmatrix}
\boldsymbol{\Sigma}_r & \boldsymbol{\Sigma}_r^{(k)}\mathbf{U}_{r,j}^\top\\
\mathbf{U}_{r,j}\boldsymbol{\Sigma}_r^{(k)}&\mathbf{V}^{(k)}
\end{bmatrix}\right)
\end{eqnarray*}
where $\mathbf{V}^{(k)}= \sigma^{(k)2} \mathbf{I}_n+\sum_{r=1}^{R} \mathbf{Z}_r \mathbf{W}_r^{(k)} \mathbf{Z}_r^\top$. The standard results of the conditional distribution of the multivariate Gaussian distribution \cite{tong_fundamental_1990} give us that:
\begin{eqnarray*}
\boldsymbol{\mu}_{r,j}^{(k)} & = & \boldsymbol{\Sigma}_r^{(k)}\mathbf{U}_{r,j}^\top\mathbf{V}^{(k)-1}(\mathbf{Y}- \mathbf{X}\boldsymbol{\beta}^{(k)}),
\\
\boldsymbol{\Omega}_{r,j}^{(k)} & = & \boldsymbol{\Sigma}_r^{(k)} - \boldsymbol{\Sigma}_r^{(k)}\mathbf{U}_{r,j}^\top\mathbf{V}^{(k)-1}\mathbf{U}_{r,j}\boldsymbol{\Sigma}_r^{(k)} + \boldsymbol{\mu}_{r,j}^{(k)}\boldsymbol{\mu}_{r,j}^{(k)\top}.
\end{eqnarray*}

\subsubsection*{M-step}
We first maximise the function $f_{\theta^{(k)}}$  with respect to $\boldsymbol{\beta}$. If we rewrite $f_{\theta^{(k)}}(\boldsymbol{\theta})$, focussing only on the terms involving $\boldsymbol{\beta}$, and regroup the other terms in a constant $c$ we get: 
\begin{eqnarray*}
f_{\theta^{(k)}}(\boldsymbol{\beta}) & = & c-\frac{1}{2\sigma^2}\big[\|\mathbf{Y}-\mathbf{X}\boldsymbol{\beta} \|^2_2
-2\big(\mathbf{Y}-\mathbf{X}\boldsymbol{\beta}\big)^\top \sum_{r=1}^R\sum_{j=1}^{m_r} \mathbf{U}_{r,j} \boldsymbol{\mu}_{r,j}\big]-\frac{\lambda}{2\sigma^2}\boldsymbol{\beta}^\top\boldsymbol{\beta},
\end{eqnarray*}
which, by modifying the constant, can be rewritten to 
\begin{eqnarray*}
f_{\theta^{(k)}}(\boldsymbol{\beta}) & = & c-\frac{1}{2\sigma^2}\big[\|\mathbf{Y}- \sum_{r=1}^R\sum_{j=1}^{m_r} \mathbf{U}_{r,j} \boldsymbol{\mu}_{r,j}-\mathbf{X}\boldsymbol{\beta} \|^2_2
-\lambda\boldsymbol{\beta}^\top\boldsymbol{\beta}\big].
\end{eqnarray*}
We here recognise the known maximisation problem of a ridge regression for which the maximum is:
\begin{eqnarray*}
\boldsymbol{\beta}^{(k+1)} & = & (\mathbf{X}^\top \mathbf{X} + \lambda \mathbf{I}_p)^{-1}\mathbf{X}^\top (\mathbf{Y}- \sum_{r=1}^R\sum_{j=1}^{m_r} \mathbf{U}_{r,j} \boldsymbol{\mu}^{k}_{r,j} ).
\end{eqnarray*}
We now turn our attention to maximizing of $f_{\theta^{(k)}}$  with respect to $\boldsymbol{\Sigma}$. First note that $\log \big[p(\mathbf{Y} \, | \, \theta, \boldsymbol{\gamma})\big ]$ does not depend on $\boldsymbol{\Sigma}$. Moreover, $p(\boldsymbol{\gamma} \, | \, \boldsymbol{\theta}) = p(\boldsymbol{\gamma} \, | \, \boldsymbol{\Sigma})$ and $\pi(\boldsymbol{\Sigma} \, | \, \boldsymbol{\Theta})$ are conjugate and the inverse Wishart law has a closed form expression for it's mode:
\begin{eqnarray*}
\boldsymbol{\Sigma}^{(k+1)}_{r} & = & (m_r+\eta_r+q_r+1)^{-1} \Big(
\boldsymbol{\Phi}_r+\sum_{j=1}^{m_r}\boldsymbol{\Omega}_{r,j}^{(k)} \Big). \end{eqnarray*}
Finally, if we rewrite $f_{\theta^{(k)}}(\boldsymbol{\theta})$ as a function of $\sigma^2$ and replace the other parameters by their maximal value we get:
\begin{eqnarray*}
f_{\theta^{(k)}}(\boldsymbol{\sigma}) & = & c - \frac{1}{2} [p+n+2(\alpha+1)] \log(\sigma^2)-\tfrac{1}{2\sigma^2}\big[\lambda\|\boldsymbol{\beta}_{k+1}\|^2_2 +2\alpha 
\\
& & +\|\mathbf{Y}-\mathbf{X}\boldsymbol{\beta}_{k+1} \|^2_2
-2\big(\mathbf{Y}-\mathbf{X}\boldsymbol{\beta}_{k+1}\big)^\top \sum_{r=1}^R\sum_{j=1}^{m_r} \mathbf{U}_{r,j} \boldsymbol{\mu}^{(k)}_{r,j}+\sum_{r,r'=1}^R\sum_{j,j'=1}^{m_r}\Tr( \mathbf{U}_{r',j'}^\top \mathbf{U}_{r,j}  \boldsymbol{\Omega}_{r,j}^{(k)})\big].
\end{eqnarray*}
It is maximized by
\begin{eqnarray*}
\boldsymbol{\sigma}^{(k+1)} & = & [p+n+2(\alpha+1)]^{-1} \big\{
\lambda\|\boldsymbol{\beta}_{k+1}\|^2_2 + 2\alpha  +\|\mathbf{Y}-\mathbf{X}\boldsymbol{\beta}_{k+1} \|^2_2 
\\
& & \qquad \qquad - 2 \big(\mathbf{Y}-\mathbf{X}\boldsymbol{\beta}_{k+1}\big)^\top \sum_{r=1}^R\sum_{j=1}^{m_r} \mathbf{U}_{r,j} \boldsymbol{\mu}^{(k)}_{r,j}+\sum_{r,r'=1}^R\sum_{j,j'=1}^{m_r}\Tr( \mathbf{U}_{r',j'}^\top \mathbf{U}_{r,j}  \boldsymbol{\Omega}_{r,j}^{(k)})
\big\}.
\end{eqnarray*}
\mbox{ }
\\
\subsubsection*{Reparameterisation of the inverse-Wishart prior}
An interesting thing to note is that the update rules for maximising the likelihood can be derived following the same procedure:
\begin{eqnarray*}
\boldsymbol{\beta}^{(k+1)} & = & (\mathbf{X}^\top \mathbf{X})^{-1}\mathbf{X}^\top \Big(\mathbf{Y}- \sum_{r=1}^R\sum_{j=1}^{m_r} \mathbf{U}_{r,j} \boldsymbol{\mu}_{r,j}  \Big)
\\
\boldsymbol{\Sigma}^{(k+1)}_{r} & = & \frac{1}{m_r}\sum_{j=1}^{m_r}\boldsymbol{\Omega}_{r,j}^{(k)}
\\
\boldsymbol{\sigma}^{(k+1)} & = & (p+n)^{-1}
\Big[  \|\mathbf{Y}-\mathbf{X}\boldsymbol{\beta}_{k+1} \|^2_2
-2\big(\mathbf{Y}-\mathbf{X}\boldsymbol{\beta}_{k+1}\big)^\top \sum_{r=1}^R\sum_{j=1}^{m_r} \mathbf{U}_{r,j} \boldsymbol{\mu}^{(k)}_{r,j}
\\
& & \qquad \qquad \quad + \sum_{r,r'=1}^R\sum_{j,j'=1}^{m_r}\Tr( \mathbf{U}_{r',j'}^\top \mathbf{U}_{r,j}  \boldsymbol{\Omega}_{r,j}^{(k)})  \Big].
\end{eqnarray*}

Using the above, we can rewrite the update rule of the covariance of the random effects for the MAP estimator derived in the previous section as
\begin{eqnarray*}
\boldsymbol{\Sigma}^{(k+1)}_{r} & = & \frac{\boldsymbol{\Phi}_r+\sum_{j=1}^{m_r}\boldsymbol{\Omega}_{r,j}^{(k)}}{m_r+\eta_r+q_r+1}
\\
& = & \frac{\eta_r+q_r+1}{m_r+\eta_r+q_r+1}\frac{\boldsymbol{\Phi}_r}{\eta_r+q_r+1}+\frac{m_r}{m_r+\eta_r+q_r+1}\frac{1}{m_r}\sum_{j=1}^{m_r}\boldsymbol{\Omega}_{r,j}^{(k)}
\\
& = & b_r A_r+ (1-b_r)\frac{1}{m_r}\sum_{j=1}^{m_r}\boldsymbol{\Omega}_{r,j}^{(k)},
\end{eqnarray*}
with $A_r =(\eta_r+q_r+1)^{-1} \boldsymbol{\Phi}_r$ and $b_r = (\eta_r+q_r+1) ( m_r+\eta_r+q_r+1)^{-1}$. We now have a simple link between the frequentist maximum likelihood EM update rule and the inverse-Wishart one. $A_r$ is the matrix towards which the regularised solution is pulled, whereas $b_r$ represents the strength of the regularisation. If $b_r = 1$ then the data is not used and the covariance is set to $A_r$. If $b_r = 0$ then the update rule becomes the one of the unconstrained maximum likelihood problem.

\subsubsection{Hessian matrices of the log-prior and log-likelihood }
In this section we derive the Hessian matrix of $\log[L(\boldsymbol{\theta};\mathbf{Y},\boldsymbol{\Theta})] = \log[f(\mathbf{Y} \, | \,\boldsymbol{\theta})] + \log[(\pi(\boldsymbol{\theta} \, | \, \boldsymbol{\Theta})]$ with respect to the parameters $\boldsymbol{\theta}$. Exploiting the linearity of the Hessian, we first derive the Hessian of $\log[f(\mathbf{Y} \, | \,\boldsymbol{\theta})]$ and then that of the log-priors $\log[(\pi(\boldsymbol{\theta} \, | \, \boldsymbol{\Theta})]$.

\subsubsection*{Hessian matrix of the log-likelihood }
In this section we denote by $g$ the log-likelihood, that is given by $g(\boldsymbol{\theta})=\log[f(\mathbf{Y} \, | \,\boldsymbol{\theta})] = c - \frac{1}{2}\log[\text{det}(\mathbf{V})]-\frac{1}{2}(\mathbf{Y}-\mathbf{X}\boldsymbol{\beta})\mathbf{V}^{-1}(\mathbf{Y}-\mathbf{X}\boldsymbol{\beta})^\top$, where $\mathbf{V} = \sigma^2\mathbf{I}_n+\sum_{r=1}^{R} \mathbf{Z}_r \mathbf{W}_r \mathbf{Z}_r^\top$ is a function of both the error variance $\sigma^2$ and the covariance matrices of the random effects $\boldsymbol{\Sigma}_r$. The first order derivatives of $g$ are:
\begin{eqnarray*}
\frac{d g}{d\boldsymbol{\boldsymbol{\beta}}}(\boldsymbol{\theta}) & = & \mathbf{X}\mathbf{V}^{-1}(\mathbf{Y}-\mathbf{X}\boldsymbol{\beta})^\top,
\\
\frac{d g}{du}(\boldsymbol{\theta}) & = & -\frac{1}{2}\text{Tr}\big(\mathbf{V}^{-1}\frac{d \mathbf{V}}{du}(\boldsymbol{\theta})\big)+\frac{1}{2}(\mathbf{Y}-\mathbf{X}\boldsymbol{\beta})\mathbf{V}^{-1}\frac{d \mathbf{V}}{du}(\boldsymbol{\theta})\mathbf{V}^{-1}(\mathbf{Y}-\mathbf{X}\boldsymbol{\beta})^\top,
\end{eqnarray*}
where $u$ denotes any random effect parameter or the error variance. The second order derivatives of $g$ are:
\begin{eqnarray*}
\frac{d^2 g}{d^2\boldsymbol{\boldsymbol{\beta}}}(\boldsymbol{\theta}) & = & \mathbf{X}\mathbf{V}^{-1}\mathbf{X}^\top,
\\
\frac{d^2 g}{d\boldsymbol{\beta}du}(\boldsymbol{\theta}) &= & -\mathbf{X}\mathbf{V}^{-1}\frac{d \mathbf{V}}{du}(\boldsymbol{\theta})\mathbf{V}^{-1}(\mathbf{Y}-\mathbf{X}\boldsymbol{\beta})^\top,
\\
\frac{d^2 g}{du_1du_2}(\boldsymbol{\theta}) & = & \frac{1}{2}\text{Tr}\big(\mathbf{V}^{-1}\frac{d \mathbf{V}}{du_1}(\boldsymbol{\theta})\mathbf{V}^{-1}\frac{d \mathbf{V}}{du_2}(\boldsymbol{\theta})+\mathbf{V}^{-1}\frac{d^2 \mathbf{V}}{du_1du_2}(\boldsymbol{\theta})\big)
\\
& & -(\mathbf{Y}-\mathbf{X}\boldsymbol{\beta})\mathbf{V}^{-1}\frac{d \mathbf{V}}{du_1}(\boldsymbol{\theta})\mathbf{V}^{-1}\frac{d \mathbf{V}}{du_2}(\boldsymbol{\theta})\mathbf{V}^{-1}(\mathbf{Y}-\mathbf{X}\boldsymbol{\beta})^\top
\\
& & +\frac{1}{2}(\mathbf{Y}-\mathbf{X}\boldsymbol{\beta})\mathbf{V}^{-1}\frac{d^2 \mathbf{V}}{du_1du_2}(\boldsymbol{\theta})\mathbf{V}^{-1}(\mathbf{Y}-\mathbf{X}\boldsymbol{\beta})^\top.
\end{eqnarray*}

\noindent Using the fact that the second order derivative $\frac{d^2 \mathbf{V}}{du_1du_2}(\boldsymbol{\theta}) = 0$ for all $u_1,u_2$ and $\frac{d \mathbf{V}}{d\sigma^2}(\boldsymbol{\theta}) = I_n$, $\frac{d \mathbf{V}}{d\boldsymbol{\Sigma}_{r,i,j}}(\boldsymbol{\theta}) = \mathbf{Z}_{r}\mathbf{E}_{r,i,j}\mathbf{Z}_{r}^\top$ and where $\mathbf{E}_{i,j}$ is the $q_r\times q_r$ 0-matrix with ones 1 at position $(i,j)$ and $(j,i)$ if $i\neq j$ and a single one at $(i,i)$ if $i=j$:

\begin{eqnarray*} 
\frac{d^2 g}{d\boldsymbol{\beta}d\sigma^2}(\boldsymbol{\theta}) & = & -\mathbf{X}\mathbf{V}^{-2}(\mathbf{Y}-\mathbf{X}\boldsymbol{\beta})^\top,
\\
\frac{d^2 g}{d\boldsymbol{\beta}d\boldsymbol{\Sigma}_{r,i,j}}(\boldsymbol{\theta}) & =& -\mathbf{X}\mathbf{V}^{-1}\mathbf{Z}_{r}\mathbf{E}_{r,i,j}\mathbf{Z}_{r}^\top\mathbf{V}^{-1}(\mathbf{Y}-\mathbf{X}\boldsymbol{\beta})^\top,
\\
\frac{d^2 g}{d^2\sigma^2}(\boldsymbol{\theta})& = & \frac{1}{2}\text{Tr}\big(\mathbf{V}^{-2}\big)-(\mathbf{Y}-\mathbf{X}\boldsymbol{\beta})\mathbf{V}^{-3}(\mathbf{Y}-\mathbf{X}\boldsymbol{\beta})^\top,
\\
\frac{d^2 g}{d\boldsymbol{\Sigma}_{r,i,j}d\boldsymbol{\Sigma}_{r',i',j'}}(\boldsymbol{\theta}) & = & \frac{1}{2}\text{Tr}\big(\mathbf{V}^{-1}\mathbf{Z}_{r}\mathbf{E}_{r,i,j}\mathbf{Z}_{r}^\top\mathbf{V}^{-1}\mathbf{Z}_{r'}\mathbf{E}_{r',i',j'}\mathbf{Z}_{r'}^\top\big)
\\
& & -(\mathbf{Y}-\mathbf{X}\boldsymbol{\beta})\mathbf{V}^{-1}\mathbf{Z}_{r}\mathbf{E}_{r,i,j}\mathbf{Z}_{r}^\top\mathbf{V}^{-1}\mathbf{Z}_{r'}\mathbf{E}_{r',i',j'}\mathbf{Z}_{r'}^\top\mathbf{V}^{-1}(\mathbf{Y}-\mathbf{X}\boldsymbol{\beta})^\top,
\\
\frac{d^2 g}{d\boldsymbol{\Sigma}_{r,i,j}d\sigma}(\boldsymbol{\theta}) & = & \frac{1}{2}\text{Tr}\big(\mathbf{V}^{-2}\mathbf{Z}_{r}\mathbf{E}_{r,i,j}\mathbf{Z}_{r}^\top\big)-(\mathbf{Y}-\mathbf{X}\boldsymbol{\beta})\mathbf{V}^{-1}\mathbf{Z}_{r}\mathbf{E}_{r,i,j}\mathbf{Z}_{r}^\top\mathbf{V}^{-2}(\mathbf{Y}-\mathbf{X}\boldsymbol{\beta})^\top.
\end{eqnarray*}
\mbox{ }
\\

\subsubsection*{Hessian matrix of the log-prior }
We now focus on the log-prior :
\begin{eqnarray*}
 \log \big[ \pi(\boldsymbol{\theta} \, | \, \boldsymbol{\Theta})\big] & =& c-\frac{p}{2}\log(\sigma^2)-\frac{\lambda}{2\sigma^2}\boldsymbol{\beta}^\top\boldsymbol{\beta}-(\alpha+1)\log(\sigma^2) -\frac{\alpha}{\sigma^2} 
\\
&  & -\sum_{r=1}^R\frac{\eta_r+q_r+1}{2}\log(|\boldsymbol{\Sigma}_r|)-\frac{1}{2} \Tr( \boldsymbol{\Sigma}_r^{-1}\boldsymbol{\Phi}_r).
\end{eqnarray*}
Note that only $\sigma^2$ and $\boldsymbol{\beta}$ interact, which simplifies the Hessian. Only the following blocks are non zero:
\begin{eqnarray*} 
\frac{d^2 g}{d^2\boldsymbol{\boldsymbol{\beta}}}(\boldsymbol{\theta}) & = & -\frac{\lambda}{\sigma^2}\mathbf{I}_p,
\\
\frac{d^2 g}{d\boldsymbol{\beta}d\sigma^2}(\boldsymbol{\theta}) & = & \frac{\lambda}{\sigma^4}\boldsymbol{\beta}
\\
\frac{d^2 g}{d^2\sigma^2}(\boldsymbol{\theta})& = &\frac{1}{\sigma^6}[(\frac{p}{2}+\alpha+1)\sigma^2-(\lambda\boldsymbol{\beta}^\top\boldsymbol{\beta}+2\alpha)],
\\
\frac{d^2 g}{d\boldsymbol{\Sigma}_{r,i,j}d\boldsymbol{\Sigma}_{r,i',j'}}(\boldsymbol{\theta}) & = & \frac{\eta_r+q_r+1}{2}\text{Tr}\big(\boldsymbol{\Sigma}_{r}^{-1}\mathbf{E}_{r,i,j}\boldsymbol{\Sigma}_{r}^{-1}\mathbf{E}_{r,i',j'}\big)
\\
& & -\text{Tr}\big(\boldsymbol{\Sigma}_{r}^{-1}\mathbf{E}_{r,i,j}\boldsymbol{\Sigma}_{r}^{-1}\mathbf{E}_{r,i',j'}\boldsymbol{\Sigma}_{r}^{-1}\boldsymbol{\Phi}_r\big),
\end{eqnarray*}
\clearpage
\subsection{Supplementary figures}
\begin{figure}[h!]
    \centering
    \includegraphics[scale = 0.15]{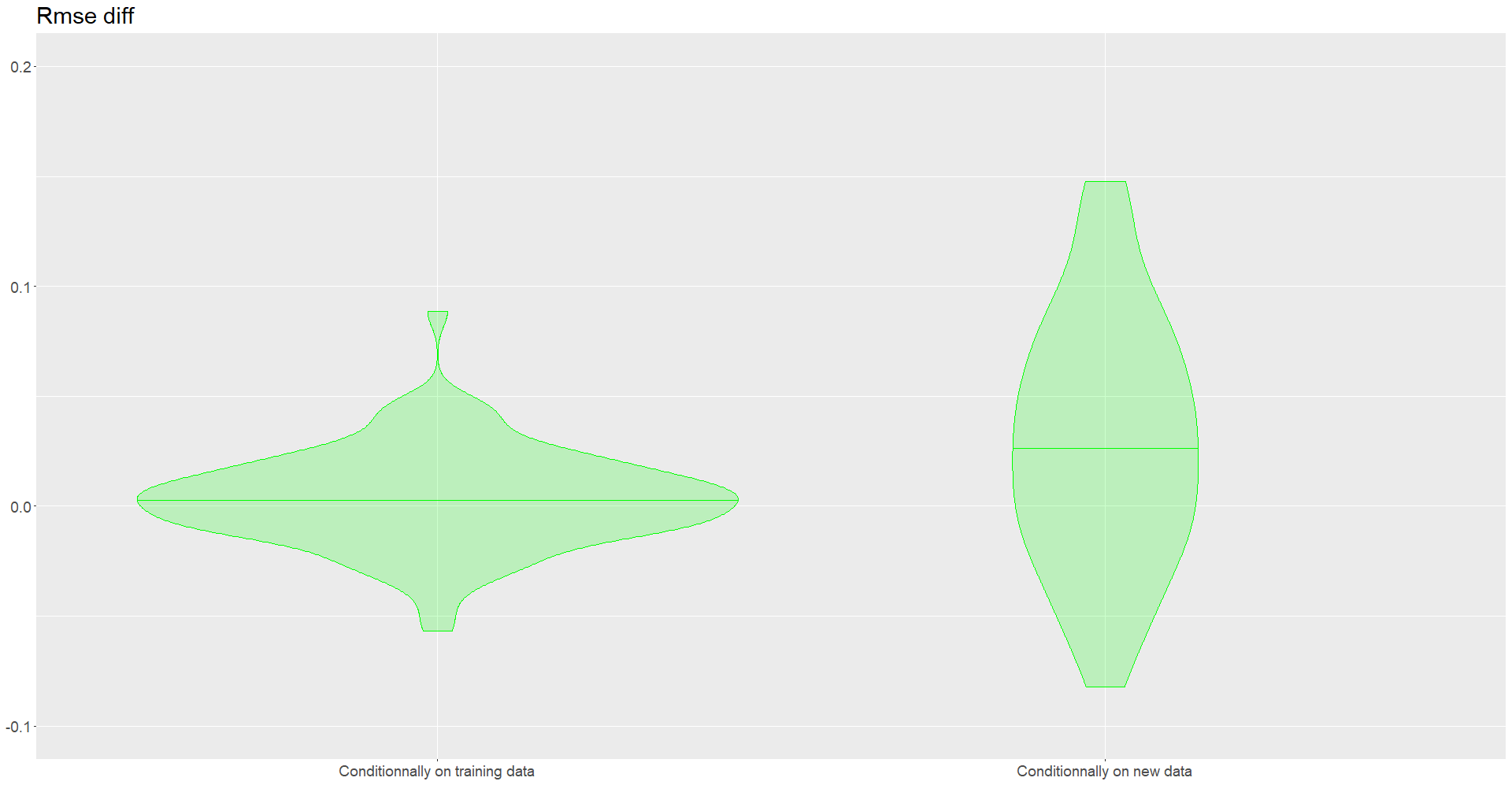}
\caption{Violinplot of the RMSE difference $\text{RMSE}(\hat{\boldsymbol{\theta}}_{\text{ML}})-\text{RMSE}(\hat{\boldsymbol{\theta}}_{\text{EB}})$ over the 150 experiments. }
    \label{fig:Rmse}
\end{figure}

\subsubsection{Speed comparison to INLA}

\subsubsection{Fixed effect hyperparameter estimation and high-dimensional fixed effects.}

\begin{figure}[h!]
    \centering
    \includegraphics[scale = 0.25]{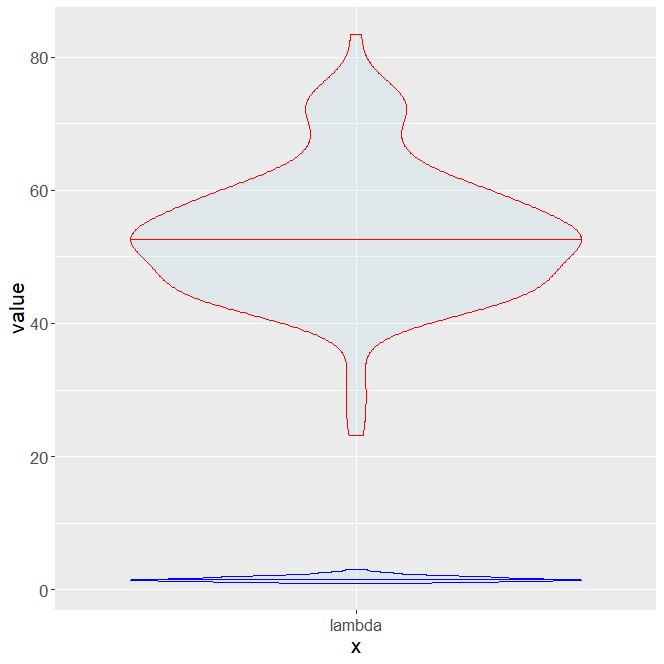}
    \caption{Violinplot of the estimated fixed effect hyperparameters when the fixed effects are respectively high-dimensional ($p=300$ and $n=240$ ) in red, and low-dimensional ($p=3$ and $n=240$ ) in blue. }
    \label{fig:highLow}
\end{figure}

Figure \ref{fig:highLow} presents a comparison of the estimated $\lambda$ in a low-dimensional fixed effect setup against that from the high-dimensional scenario. It is evident that the estimated $\lambda$ is substantially smaller in the low-dimensional case, confirming our method's accurate estimation of the fixed effect hyperparameter according to the specific scenario.

\end{document}